\documentclass[superscriptaddress,twocolumn,aps,pra,preprintnumbers,notitlepage]{revtex4-1}
\usepackage{amsmath,amssymb,esint}
\usepackage{lipsum} 
\usepackage[latin9]{inputenc}
\usepackage{graphicx}
\usepackage{dcolumn}
\usepackage{bbold}
\usepackage{bm}
\usepackage[usenames,table]{xcolor}
\usepackage[colorlinks=true,linktocpage,bookmarks=true,citecolor=blue,linkcolor=blue,urlcolor=blue,hyperfootnotes=true,pageanchor=true]{hyperref}
\usepackage[caption=false,labelformat=simple]{subfig}
\usepackage{accents}
\usepackage[export]{adjustbox}
\usepackage{mathtools}
\usepackage{float}
\usepackage{mathrsfs}
\usepackage{pifont}

\usepackage{adjustbox}
\usepackage{multirow}

\usepackage{soul} 

\usepackage{tikz}

\definecolor{lime}{HTML}{A6CE39}
\DeclareRobustCommand{\orcidicon}{%
	\begin{tikzpicture}
	\draw[lime, fill=lime] (0,0) 
	circle [radius=0.15] 
	node[white] {{\fontfamily{qag}\selectfont \tiny ID}};
	\draw[white, fill=white] (-0.0625,0.095) 
	circle [radius=0.007];
	\end{tikzpicture}
	\hspace{-3mm}
}

\foreach \x in {A, ..., Z}{\expandafter\xdef\csname orcid\x\endcsname{\noexpand\href{https://orcid.org/\csname orcidauthor\x\endcsname}{\noexpand\orcidicon}}
}

\begin{document}

\title{Slave-rotor theory of correlated altermagnets on the Lieb lattice}

\author{Vanuildo S. de Carvalho\orcidA{}}
\email{vanuildo\_carvalho@ufg.br}
\affiliation{Instituto de F\'{\i}sica, Universidade Federal de Goi\'as, 74.690-900, Goi\^ania-GO,
Brazil}
\author{Hermann Freire\orcidB{}}
\affiliation{Instituto de F\'{\i}sica, Universidade Federal de Goi\'as, 74.690-900, Goi\^ania-GO,
Brazil}
\author{Rodrigo G. Pereira\orcidC{}}
\affiliation{International Institute of Physics and Departamento de F\'isica Te\'orica e Experimental, Universidade Federal do Rio Grande do Norte, Campus Universit\'ario, Lagoa Nova,  Natal, RN, 59078-970, Brazil}

\begin{abstract}
We investigate the metal-insulator transition driven by the onsite repulsive interaction $U$ in an altermagnetic Hubbard model defined on a Lieb lattice. Using the slave-rotor approach at half-filling, we find that the system exhibits a cascade of interaction-driven phase transitions. As $U$ increases, the  system evolves from a normal metal to an altermagnetic metal, then to an altermagnetic insulator, and eventually to an altermagnetic Mott insulator characterized by the complete suppression of the quasiparticle weight. These phases are supported by the calculation of the electronic spectral function, which features spin-split bands in both the metallic and insulating regimes. However, the spin splitting becomes substantially suppressed in the Mott-insulating phase. Our results  suggest  that the observation of spin splitting in the spectral function of $d$-wave altermagnets with a Lieb-lattice-like structure may be limited to the weak-to-moderate correlation regime.  
\end{abstract}

\date{\today}

\maketitle

\section{Introduction}\label{SecI}
\noindent 

The standard theory of collinear magnetic orders \cite{Landau-SP(1980),Ashcroft-SSP(1976)}, which describes ferromagnetism and antiferromagnetism as the main motifs in magnetic systems, has recently undergone a major revision with the realization that it does not encompass the novel altermagnetic ordering \cite{Kusunose-JPSJ(2019),Jungwirth-PRX(2022a),Jungwirth-PRX(2022b),Turek-PRB(2022)}. This state refers to the absence of net macroscopic magnetization due to the combination of time-reversal symmetry with a point-group operation, representing either reflection or rotation \cite{Jungwirth-arXiv(2024),Smejkal-New(2025),Jungwirth-N(2026)}. As a result, the order parameter describing altermagnets acquires an even-parity angular momentum symmetry (e.g., $d$, $g$, or $i$ wave), which in the simplest case implies the magnetic-counterpart realization of a $d$-wave high-$T_c$ superconductor. Additionally, altermagnets are characterized by   nonuniform spin-split bands, which become degenerate at nodal manifolds dictated  by the point-group symmetry \cite{Fernandes-PRB(2024),Agterberg-PRB(2024)}. Most importantly, these features are expected to have an immediate impact on the development of next-generation spintronic devices.

On the experimental front, altermagnetism was first confirmed in the three-dimensional (3D) compounds $\mathrm{CrSb}$ \cite{Reimers-NC(2024),Ding-PRL(2024),Zeng-AS(2024),Brink-CP(2025)} and $\mathrm{MnTe}$ \cite{Kriegner-PRL(2023),Lee-PRL(2024),Osumi-PRB(2024),Jungwirth-N(2024),Amin-N(2024)}. Indeed, angle-resolved photoemission spectroscopy (ARPES) unveiled in these materials the existence of a $g$-wave altermagnetic phase with a nonuniform spin-splitting reaching the electron-volt figure in some parts of the Brillouin zone (BZ) \cite{Osumi-PRB(2024),Yang-NC(2025)}. Theoretical investigations involving \emph{ab initio} methods \cite{Facio-MTP(2023)}, group theory analysis of magnetic structures \cite{Xiao-PRX(2024)}, and artificial-intelligence algorithms \cite{Gao-NSR(2025)} have subsequently increased the number of altermagnetic candidates to more than 100 compounds, revealing the ubiquity of this phenomenon.

\begin{figure}[t]
\centering
\includegraphics[width=0.8\linewidth,valign=t]{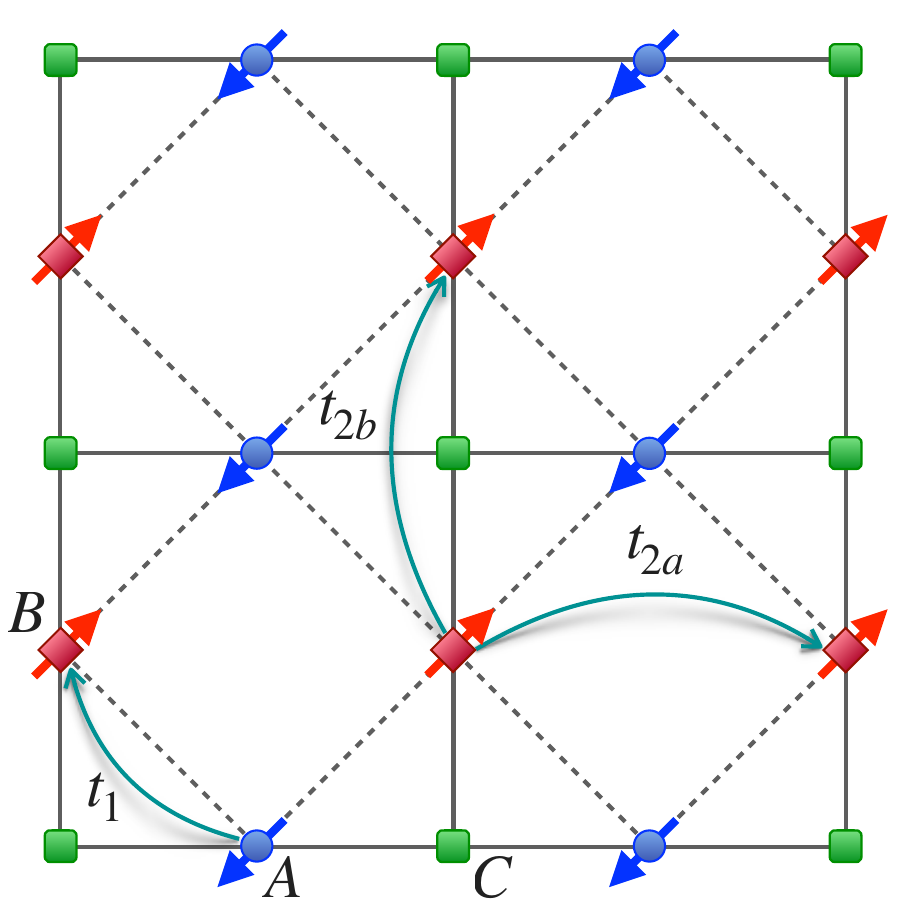}
\caption{Schematic representation of the Lieb-lattice structure   in quasi-2D altermagnets with a $T_2\mathrm{O}$ structure, where the magnetic $T$ (nonmagnetic $\mathrm{O}$) atoms occupy the $A$ and $B$ ($C$) sublattice sites. The altermagnetic configuration is indicated by   blue and red arrows, representing opposite magnetic moments. The electrons described by the model Hamiltonian \eqref{HubbHam}   move between the $A$ and $B$ sublattices with hopping amplitudes $t_1$ and $t_{2a/2b}$ for nearest-neighbor (NN) and next-nearest-neighbor (NNN) sites, respectively. }\label{Fig1}
\end{figure}

In this regard, the long sought-after $d$-wave altermagnetic state was first predicted to emerge in the metallic system $\mathrm{RuO_2}$ \cite{Kunes-PRB(2019),Smejkal-SA(2020)}. Compared to materials with $g$-wave symmetry, $d$-wave altermagnets have an advantage for the production of spin-polarized currents \cite{Jiang-NP(2025)}. While some experimental probes initially pointed to $d$-wave altermagnetism in $\mathrm{RuO_2}$ \cite{Feng-NE(2022),Bai-PRL(2022),Karube-PRL(2022),Bose-NE(2022),Bai-PRL(2023)}, the status of this material is currently under debate after $\mu$SR and neutron diffraction experiments found no sign of magnetic order in its phase diagram \cite{Hiraishi-PRL(2024),Moser-npjS(2024)}. Nevertheless, spin-resolved ARPES recently identified $d$-wave altermagnetic order in the insulator $\mathrm{La_2O_3Mn_2Se_2}$ \cite{PRM-Wei(2025)} and in the   metallic systems $\mathrm{KV_2Se_2O}$ \cite{Jiang-NP(2025)} and $\mathrm{Rb_{1 - \delta}V_2Te_2O}$ \cite{Zhang-NP(2025)}. Importantly, these altermagnetic candidates are classified as quasi-two-dimensional (quasi-2D) oxychalcogenide materials, formed by intercalated $T_2\mathrm{O}$ layers with transition metal ($T$) and oxygen ($\mathrm{O}$) atoms organized in a geometric structure known as Lieb or ``anti-$\mathrm{CuO_2}$'' lattice (see Fig. \ref{Fig1}).

Moreover, recent density-functional-theory (DFT) calculations have   predicted the existence of a family of altermagnetic candidates with an underlying Lieb lattice \cite{Mazin-npgQM(2026)}. This  result  highlights the importance of investigating the phase diagram of paradigmatic correlated electronic models defined on this lattice. For this reason, the effect of strong interactions on the phase diagram of altermagnetic Lieb-lattice Hubbard models has attracted increasing attention in recent years. In fact, this question has been addressed by several methods, including, e.g., effective spin-fermion models \cite{Venderbos-PRL(2025)}, mean-field techniques \cite{Knap-PRL(2024)}, functional renormalization group \cite{Thomale-PRL(2025)}, Hartree-Fock and exact diagonalization \cite{Franz-PRL(2025)}, and also slave-particle \cite{Giuli-PRB(2025),Sobral-PRR(2025),Daghofer-PRL(2026),Zeng-PRB(2026),Dong-PRB(2026),Seufert-arXiv(2026)} approaches. Among other findings, the phase diagram has been shown to contain both metallic and band-insulating $d$-wave altermagnetic states, which may agree at least qualitatively with experimental reports of altermagnetism in oxychalcogenides.

Motivated by the possibility of a Mott transition in strongly correlated altermagnets \cite{Fernandes-PRM(2025),DelRe-PRR(2025),Park-arXiv(2026)}, in this work we perform  a slave-rotor study \cite{Georges-PRB(2002),Georges-PRB(2004)} of the phase diagram of a half-filled  Hubbard model on the Lieb lattice \cite{Venderbos-PRL(2025)}. The choice of this approach is motivated by its reliability in the description of metal-insulator transitions in  various systems \cite{Lee-PRL(2005),Balents-NP(2010),Ko-PRB(2011),Huang-PRB(2020),He-PRB(2022),Balents-PRB(2023),Sangiovanni-PRL(2024)}. We find that the increase of the Hubbard onsite interaction $U$ drives the  normal metal ($\mathrm{NM}$) phase into an altermagnetic metal ($\mathrm{A\ell M}$),  which then turns into an altermagnetic insulator ($\mathrm{A\ell I}$), followed by  an altermagnetic Mott insulator ($\mathrm{A\ell MI}$). The latter is signaled by the vanishing of the quasiparticle weight and the gap opening in the density of states. Within the $\mathrm{A\ell MI}$ phase, where electrons fractionalize into gapped chargons and  spinons, the spectral function still exhibits a small $d_{x^2-y^2}$-wave spin splitting. However, the spin splitting eventually vanishes upon further increase of the Hubbard interaction.  This result suggests that this feature of altermagnetism in Lieb-lattice compounds may be restricted to the weak-to-moderate coupling regime, thereby narrowing the class of quasi-2D altermagnetic candidates  predicted solely from a symmetry analysis of their magnetic structures \footnote{Although these findings could in principle be obtained within more conventional slave-particle methods, they were not anticipated by the works in Refs. \cite{Giuli-PRB(2025),Sobral-PRR(2025),Daghofer-PRL(2026),Zeng-PRB(2026),Dong-PRB(2026),Seufert-arXiv(2026)}, which rely on the application of these other approaches to study the effect of strong electronic correlations in altermagnetic models.}.

The remainder of this work is structured as follows. In Sec. \ref{SecII}, we introduce the Hubbard model used to describe the altermagnetic order on the Lieb lattice and then apply the slave-rotor approach to obtain the self-consistent equations governing its order parameters. Section \ref{SecIII} addresses the phase diagram of this model from the numerical solution of these equations. Then, in Sec. \ref{SecIV}, we discuss the spectroscopic features and the behavior of the density of states throughout the phases exhibited by this model. Finally, our conclusions are presented in Sec. \ref{SecV}. We leave to the Appendixes most of the technical details concerning the application of the slave-rotor theory.

\section{Model and slave-rotor theory}\label{SecII}
\noindent

We consider a half-filled system of spin-$\frac{1}{2}$ electrons hopping  on the $A$ and $B$ sites of a Lieb lattice as displayed in Fig. \ref{Fig1}. In a 2D material formed by stacked $T_2\mathrm{O}$ planes, these sites are occupied by the $T$ atoms, while the oxygen occupies the $C$ sites. The Lieb lattice lacks an inversion center; however, opposite magnetic moments on the $A$ and $B$ sublattices are related by a rotation, which can therefore give rise to altermagnetic order. The simplest model that describes  altermagnetism with a repulsive Hubbard interaction $U$ between  electrons is defined by the Hamiltonian \cite{Knap-PRL(2024),Venderbos-PRL(2025)}
\begin{align}
H = & - \sum_{\mathbf{r}, \mathbf{r}', \sigma} (t_{\mathbf{r}, \mathbf{r}'} c^{\dagger}_{\mathbf{r}, \sigma} c_{\mathbf{r}', \sigma} + \mathrm{H. c.}) - \mu \sum_{\mathbf{r}, \sigma} c^{\dagger}_{\mathbf{r}, \sigma} c_{\mathbf{r}, \sigma}, \nonumber \\
& + U \sum_{\mathbf{r}} \bigg(n_{\mathbf{r}, \uparrow} - \frac{1}{2}\bigg)\bigg(n_{\mathbf{r}, \downarrow} - \frac{1}{2}\bigg), \label{HubbHam}
\end{align}
where the vector $\mathbf{r} \equiv (\boldsymbol{r}, s)$ comprises the unit cell ($\boldsymbol{r}$) and sublattice ($s \in \{A, B \}$) coordinates   with electron  creation (annihilation) operator $c^{\dagger}_{\mathbf{r}, \sigma}$ ($ c_{\mathbf{r}, \sigma}$), number operator $n_{\mathbf{r}, \sigma} \equiv c^{\dagger}_{\mathbf{r}, \sigma}c_{\mathbf{r}, \sigma}$, and chemical potential $\mu$. In addition,  $t_{\mathbf{r}, \mathbf{r}'}$ denotes nearest-neighbor (NN) hopping $t_1$ and next-nearest-neighbor (NNN) hoppings $t_{2a}$ and $t_{2b}$. As  depicted in Fig. \ref{Fig1}, the anisotropy in the latter stems from  different crystallographic environments of the $A$ and $B$ sublattices \cite{Venderbos-PRL(2025)}.

To investigate  model \eqref{HubbHam} within  slave-rotor theory \cite{Georges-PRB(2002),Georges-PRB(2004)}, we start by  introducing the mapping $c_{\mathbf{r}, \sigma} = e^{i \theta_{\mathbf{r}}} \psi_{\mathbf{r}, \sigma}$. Here, $\psi_{\mathbf{r}, \sigma}$ describes spin-$\frac{1}{2}$ fermionic spinons, while the spinless $O(2)$ bosonic field $\theta_j$  is associated with chargons. The enlarged Hilbert space of   spinons and chargons in this  fractionalization scheme is constrained by $\mathscr{L}_{\mathbf{r}} = \sum_{\sigma} \left( \psi^{\dagger}_{\mathbf{r}, \sigma}\psi_{\mathbf{r}, \sigma} - \frac{1}{2} \right)$, where $\mathscr{L}_{\mathbf{r}}$ is the angular momentum operator conjugated to $\theta_{\mathbf{r}}$, which obeys the canonical commutation relation $[\theta_{\mathbf{r}}, \mathscr{L}_{\mathbf{r}^\prime}] = i \delta_{\mathbf{r}, \mathbf{r}^\prime}$. Within this theory, $\mathscr{L}_{\mathbf{r}}$ plays the role of a charge quantum number, while $e^{i \theta_{\mathbf{r}}}$ behaves as its lowering operator.

In the path-integral formulation of the slave-rotor method, we first have to disentangle the charge and spin degrees of freedom related to the Hubbard interaction. This can be achieved by means of the identity \cite{Castellani-PLA(1979),Macedo-EL(1987),Macedo-PRB(1991)}
\begin{align}\label{HubbParam}
n_{\mathbf{r}, \uparrow}n_{\mathbf{r}, \downarrow} = & - \frac{1}{2}(b_0 - 1)(n_{\mathbf{r}, \uparrow} + n_{\mathbf{r}, \downarrow}) + \frac{b_0}{4}(n_{\mathbf{r}, \uparrow} +n_{\mathbf{r}, \downarrow})^2 \nonumber \\
& - \frac{b_z}{4}(n_{\mathbf{r}, \uparrow} - n_{\mathbf{r}, \downarrow})^2,
\end{align}
where $b_0 + b_z = 2$. However, among the infinite ways that one can parametrize Eq. \eqref{HubbParam}, it has been demonstrated that the only physical solution consistent with both the Pauli exclusion principle and the spin-rotation symmetry turns out to be $b_0 = b_z = 1$ \cite{Macedo-PRB(1991),Hamann-PRL(1969),Schrieffer-PRL(1969),Schulz-PRL(1990)}. Using this result into Eq. \eqref{HubbParam}, we find that the slave-rotor Lagrangian for the altermagnetic Hubbard model \eqref{HubbHam} becomes
\begin{align}\label{SRotorLag}
\mathcal{L} & = \sum_{\mathbf{r}} \bigg[- i \mathscr{L}_{\mathbf{r}} \partial_{\tau} \theta_{\mathbf{r}} + \sum_{\sigma}\bar{\psi}_{\mathbf{r}, \sigma} (\partial_{\tau} - \mu - h_{\mathbf{r}})\psi_{\mathbf{r}, \sigma} \nonumber \\
& + h_{\mathbf{r}}(\mathscr{L}_{\mathbf{r}} + 1) + \frac{U}{4} \mathscr{L}^2_{\mathbf{r}} - \frac{U}{4} \bigg(\sum_{\sigma, \sigma^{\prime}}\bar{\psi}_{\mathbf{r}, \sigma} \sigma^z_{\sigma, \sigma^\prime}\psi_{\mathbf{r}, \sigma^\prime} \bigg)^2\bigg] \nonumber \\
& - \sum_{\mathbf{r}, \mathbf{r}^{\prime}, \sigma} \big[ t_{\mathbf{r}, \mathbf{r}'} e^{- i(\theta_{\mathbf{r}} - \theta_{\mathbf{r}'})} \bar{\psi}_{\mathbf{r}, \sigma} \psi_{\mathbf{r}', \sigma} + \mathrm{H. c.} \big]. 
\end{align}
Here, $h_{\mathbf{r}} = h_{\mathbf{r}}(\tau)$ is a Lagrange multiplier that enforces the Hilbert-space constraint. Instead of working with phase fields $\theta_{\mathbf{r}}(\tau)$, we henceforth use complex bosonic operators $X_{\mathbf{r}}(\tau) \equiv e^{i \theta_{\mathbf{r}}(\tau)}$ to describe the charge degrees of freedom  \cite{Georges-PRB(2002),Georges-PRB(2004)}. To enforce the constraint $\vert X_{\mathbf{r}}(\tau) \vert^2 = 1$, we employ a second Lagrange multiplier $\lambda_{\mathbf{r}}(\tau)$. Next, we integrate out the angular momentum $\mathscr{L}_{\mathbf{r}}$  and  decouple the resulting Lagrangian for the staggered magnetization $(- 1)^s M_{\mathbf{r}} \equiv \frac{U}{2} \big\langle\sum_{\sigma, \sigma^{\prime}}\bar{\psi}_{\mathbf{r}, \sigma} \sigma^z_{\sigma, \sigma^\prime}\psi_{\mathbf{r}, \sigma^\prime} \big\rangle$ and the bond variables $Q_{X}^{\mathbf{r}, \mathbf{r}^\prime} \equiv \big\langle\sum_{\sigma, \sigma^{\prime}}\bar{\psi}_{\mathbf{r}, \sigma} \sigma^0_{\sigma, \sigma^\prime}\psi_{\mathbf{r}^\prime, \sigma^\prime} \big\rangle$ and $Q^{\mathbf{r}, \mathbf{r}^\prime}_{\psi} \equiv \big\langle \bar{X}_{\mathbf{r}} X_{\mathbf{r}^\prime} \big\rangle$. As a result, we arrive at a set of self-consistent equations described in Appendix \ref{AppA}. In these equations, the bond variables govern the renormalized hopping parameters for  chargons  and spinons.  The  equations also involve  the   condensation amplitude  $Z \equiv \langle T_{\tau}[X^s_{\mathbf{k} = \mathbf{0}}(\tau) X^{s' \dagger}_{\mathbf{k} = \mathbf{0}}(\tau)]\rangle/N_c$ \cite{Georges-PRB(2002),Georges-PRB(2004),Lee-PRL(2005),Sangiovanni-PRL(2024)}, also known as the quasiparticle weight, which obeys
\begin{equation}
\frac{U}{4 N_c} \sum_{\nu = \pm} \sum_{\mathbf{k} \neq \mathbf{0}} \frac{1}{E^{\nu}_X(\mathbf{k})} \coth\bigg[ \frac{\beta E^{\nu}_X(\mathbf{k})}{2} \bigg] = 1 - Z.
\end{equation}
Here, $\beta = 1/T$ is the reciprocal temperature, $N_c$ is the number of unit cells, and $E^{\pm}_X(\mathbf{k})$ refers to  the dispersion of the chargon bands. In the Fermi-liquid regime, we have  $0 < Z \leqslant 1$ and  chargons  are condensed, while in the Mott-insulating phase  the quasiparticle weight vanishes and chargons acquire a Mott gap $\Delta_X$.

\begin{figure}[t]
\centering
\includegraphics[width=1.0\linewidth,valign=t]{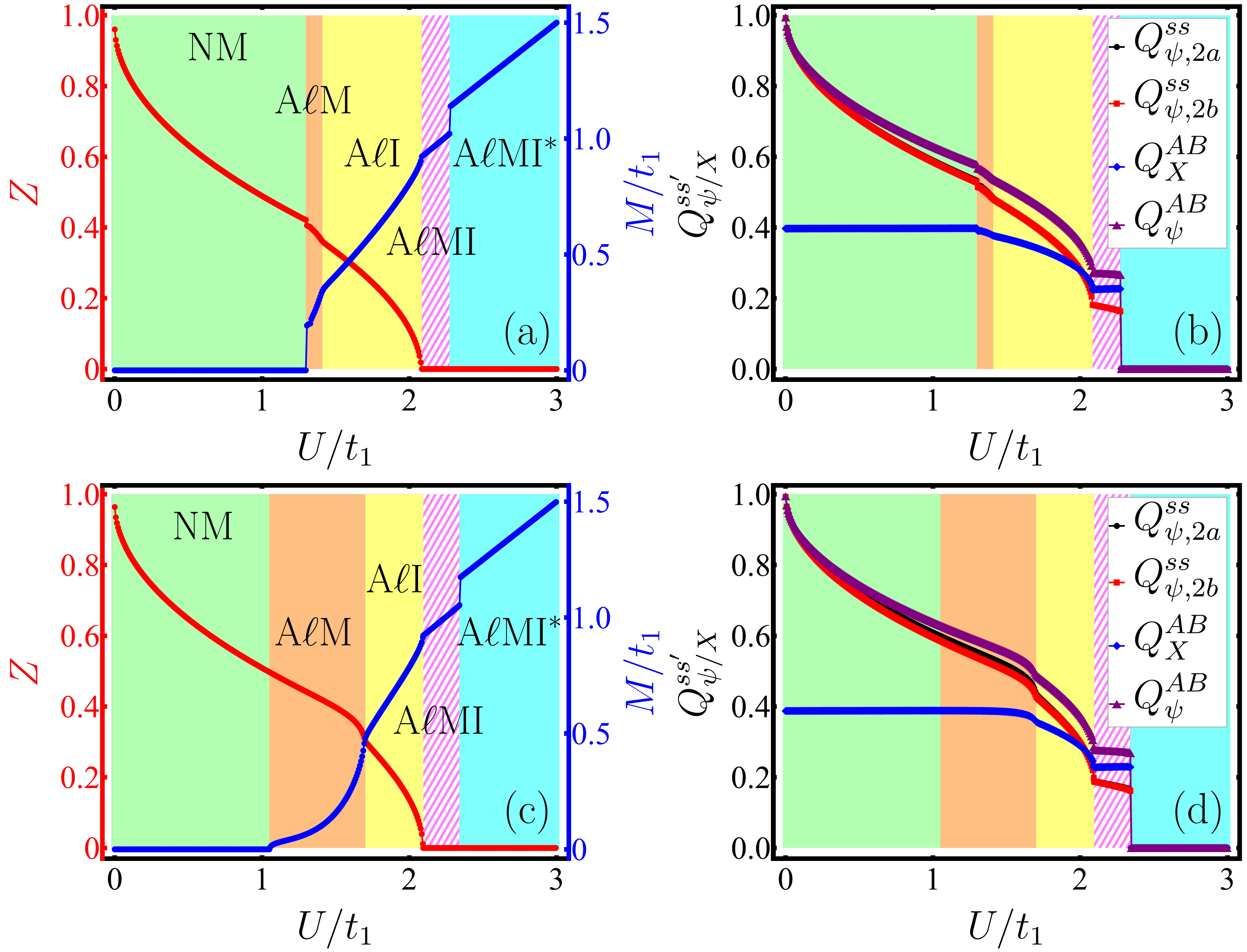}
\caption{Order parameters and phase diagram of the altermagnetic Hubbard model at half-filling. In the NNN hoppings  $t_{2a/2b} = t^{\prime}(1 \pm \delta)$, we fix   $t^{\prime} = 0.3 t_1$  and consider two values of the anisotropy parameter:  [(a) and (b)] $\delta = 0.2$; [(c) and (d)]  $\delta = 0.8$. The temperature is kept at $T = 0.01 t_1$. The altermagnetic order sets in when the staggered magnetization $M$ becomes finite. Depending on the anisotropy, this transition can be either continuous or discontinuous. As the onsite repulsion $U$ increases, the order parameters exhibit a series of nonanalyticities which signal  transitions between a  normal metal (NM), an altermagnetic metal ($\mathrm{A\ell M}$), an altermagnetic insulator ($\mathrm{A\ell I}$), and an altermagnetic Mott insulator ($\mathrm{A\ell MI}$). The latter is  characterized by a vanishing quasiparticle weight $Z$ and a finite Mott gap. The regime marked as $\mathrm{A\ell MI}^*$  features dispersionless bands and completely localized electrons due to the suppression of the bond variables $Q^{s s^\prime}_{\psi/X}$.}\label{Fig2}
\end{figure}

\section{Phase diagram}\label{SecIII}
\noindent

To solve the self-consistent equations (see Appendix \ref{AppA}), we first notice that the bond order parameters that renormalize the NNN hoppings must obey the relations $Q^{A A}_{2a, X} = Q^{B B}_{2a, X}$, $Q^{A A}_{2a, \psi} = Q^{B B}_{2a, \psi}$, $Q^{A A}_{2b, X} = Q^{B B}_{2b, X}$, and $Q^{A A}_{2b, \psi} = Q^{B B}_{2b, \psi}$. This occurs because the chemical potential is the same for   electrons in   $A$ and $B$ sublattices. After further imposing the conditions $Q^{B A}_{X} = Q^{A B}_{X}$ and $Q^{B A}_{\psi} = Q^{A B}_{\psi}$, which do not limit the range of physical states allowed by the slave-rotor equations, the self-consistent system is reduced to nine equations. The numerical solution of these equations for temperature $T = 0.01 t_1$ is shown in Fig. \ref{Fig2}. Note that all energy scales are given in terms of the NN hopping $t_1$ and we parametrize  the NNN hoppings as $t_{2a/2b} = t^{\prime}(1 \pm \delta)$ \cite{Knap-PRL(2024)}. In Fig. \ref{Fig2}, we   fix $t^{\prime} = 0.3 t_1$ and consider  weak- and strong-anisotropy regimes,  set by $\delta = 0.2$ and $0.8$, respectively. We find that, regardless of the value of $\delta$, the quasiparticle weight $Z$ is finite for weak repulsive interaction $U$. As $U$ approaches a critical value, $Z$ goes continuously to zero, indicating  a Mott transition in the model. As described in Appendix \ref{AppA}, this result is also confirmed by the development of a finite Mott gap $\Delta_X$ in the electronic spectrum when $Z \rightarrow 0$.

The results in Fig. \ref{Fig2} also  show that the staggered magnetization $M$ becomes finite as $U$ reaches a critical interaction strength, which  defines the transition from the NM to the $\mathrm{A\ell M}$. Beyond that point, the magnetization increases  monotonically, featuring a series of kinks and discontinuities. The renormalized hoppings $Q_{X}^{s, s^\prime}$ and $Q^{s, s^\prime}_{\psi}$ show similar nonanalyticities. Importantly, the solutions of the self-consistent equations   show that $M$, $Q_{X}^{s, s^\prime}$, and $Q^{s, s^\prime}_{\psi}$ are finite in the region where $Z \rightarrow 0$, which implies that the Mott insulator possesses altermagnetic correlations.

\begin{figure}[t]
\centering
\includegraphics[width=1.0\linewidth,valign=t]{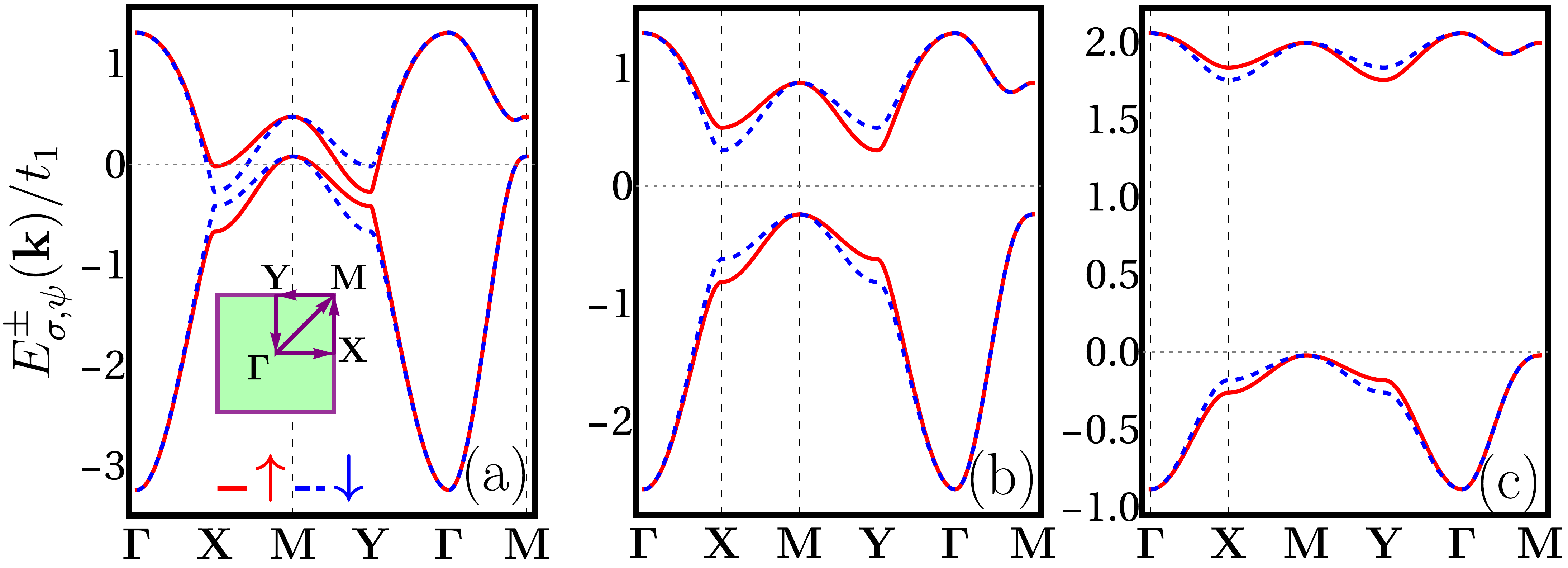} 
\caption{Spinon dispersion along a path with high-symmetry points of the BZ [see inset in (a)] for Hubbard interaction given by (a) $U \approx 1.31 t_1$, (b) $U \approx 1.71 t_1$, and (c) $U \approx 2.25 t_1$. These values of interaction correspond, respectively, to the $\mathrm{A\ell M}$, $\mathrm{A\ell I}$, and $\mathrm{A\ell MI}$ phases. Here we set $t^{\prime} = 0.3 t_1$, $\delta = 0.2$, and $T = 0.01 t_1$. The red and blue curves represent the dispersions for spin-up and spin-down spinon bands, respectively. The $\mathrm{A\ell M}$ exhibits spinon Fermi surfaces, which are gapped out in the insulating phases, as a result of the opening of either a (b) band or (c) Mott gap. All spinon dispersions shown here exhibit a  $d_{x^2 - y^2}$-wave spin splitting characteristic of altermagnetism. }\label{Fig3}
\end{figure}

To add further support to this finding, let us analyze the spinon spectrum in different regimes. Figure \ref{Fig3} shows the spinon dispersion for values of $U$ in the $\mathrm{A\ell M}$, $\mathrm{A\ell I}$, and $\mathrm{A\ell MI}$ phases, in which the order parameters $M$, $Q_{X}^{s, s^\prime}$, and $Q^{s, s^\prime}_{\psi}$ are all finite. In contrast with the NM, in which the bands are spin degenerate, in the $\mathrm{A\ell M}$ phase the spinons form Fermi surfaces  featuring $d_{x^2 - y^2}$-wave spin splittings [see Fig. \ref{Fig3}(a)] \footnote{When the anisotropy of the NNN hoppings is small, we find that the increase of the onsite repulsion $U$ induces a Lifshitz transition \cite{Lifshitz-JETP(1960a),*Lifshitz-JETP(1960b)}, in which half of the Fermi pockets located along the crystallographic axes disappear. For instance, this occurs for $\delta = 0.2$, as evidenced by the small kink observed in the order parameters within the $\mathrm{A\ell M}$ region in Figs. \ref{Fig2}(a) and \ref{Fig2}(b). The disappearance of Fermi pockets along the crystallographic axes and, therefore, the existence of a Lifshitz transition inside a metallic state with altermagnetic order was also reported in a previous work \cite{Knap-PRL(2024)}. However, Ref. \cite{Knap-PRL(2024)} obtained these findings by means of a simple Hartree-Fock theory, which relies on fixing the quasiparticle weight to $Z = 1$ for all values of the Hubbard repulsion $U$.}. As $U$ increases further, $\mathrm{A\ell M}$ transitions into the $\mathrm{A\ell I}$ phase, characterized by a finite spinon band gap and condensed chargons ($Z>0$) [see Fig. \ref{Fig3}(b)]. When $Z$ vanishes, we obtain the $\mathrm{A\ell MI}$ phase, in which both spinons and chargons are gapped and the spinon bands exhibit a finite, although small, spin splitting [see Fig. \ref{Fig3}(c)]. We also find that within the $\mathrm{A\ell MI}$ phase the chemical potential   lies slightly above the lower spinon bands. Note that the chemical potential at half-filling is not fixed in the absence of particle-hole symmetry and could be varied to be close to  the midpoint of the Mott gap.

\begin{figure}[t]
\centering
\includegraphics[width=1.0\linewidth,valign=t]{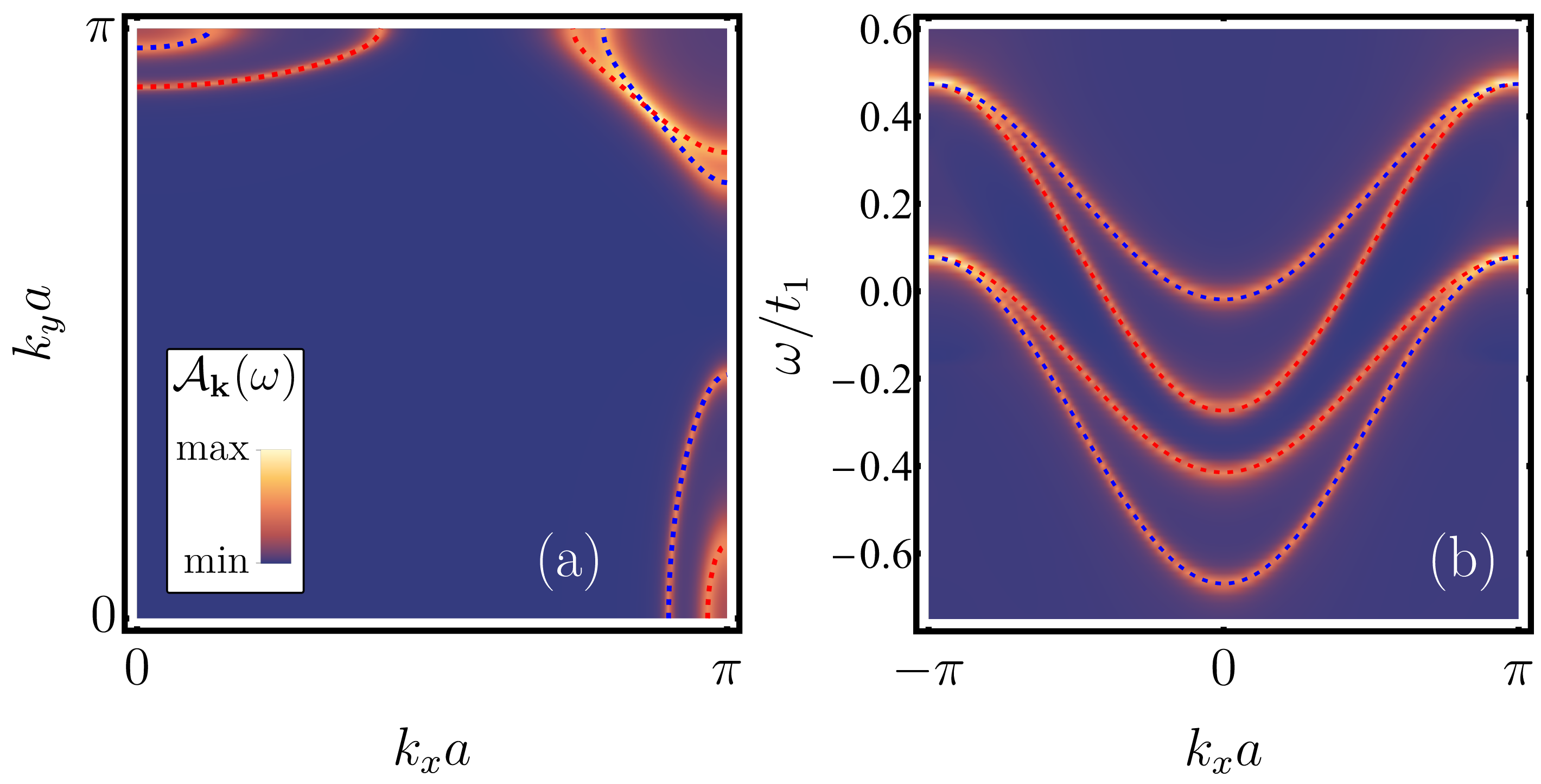} \vfil{} 
\includegraphics[width=1.0\linewidth,valign=t]{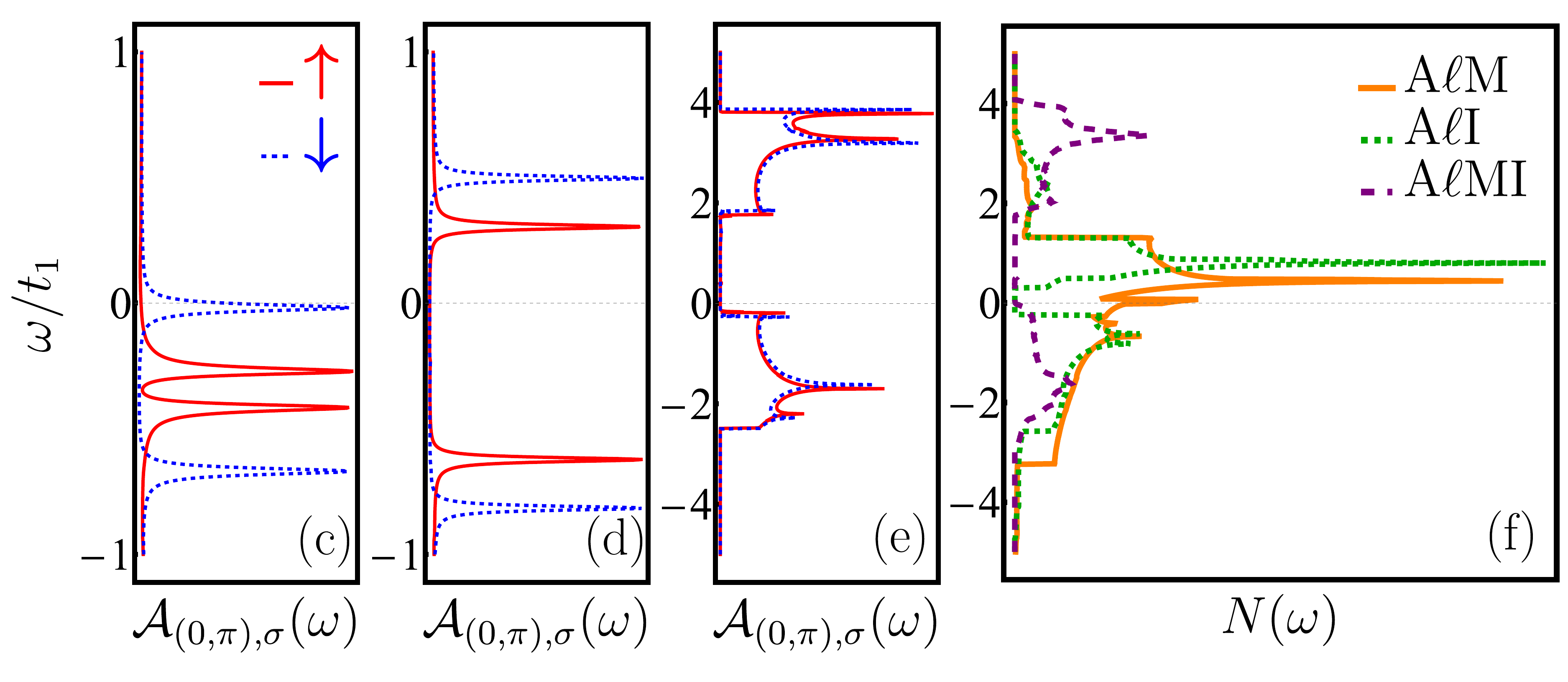}
\caption{Electron spectral function $\mathcal{A}_{\mathbf{k}}(\omega)$  for $t^\prime = 0.3 t_1$ and $\delta = 0.2$. (a)  Spectral function  $\mathcal{A}_{\mathbf{k}}(\omega=0)$ in the $\mathrm{A\ell M}$ phase; the peaks can be identified with the electron Fermi surface. The red and blue dashed lines represent, respectively, the spin-up and -down spinon Fermi surfaces.  (b) The agreement between spinon and electron spin splittings in the $\mathrm{A\ell M}$ phase is also exemplified by the dependence of $\mathcal{A}_{\mathbf{k}}(\omega)$ along the BZ cut $k_y = \pi/a$; the dashed lines represent the spinon dispersions. Both (a) and (b) are obtained for $U \approx 1.31 t_1$. We also show the energy dependence of the spectral function for $\mathbf k = (0, \pi/a)$ and values of $U$ in  three different phases: (c) $\mathrm{A\ell M}$ ($U \approx 1.31 t_1$); (d) $\mathrm{A\ell I}$ ($U \approx 1.71 t_1$); and (e) $\mathrm{A\ell MI}$ ($U \approx 2.25 t_1$). Note that the spin splitting decreases as $U$ increases. (f) The   density of states $N(\omega)$ shows a gap in the $\mathrm{A\ell I}$ and $\mathrm{A\ell MI}$ phases. }\label{Fig4}
\end{figure}

As  shown in Fig. \ref{Fig2}, the order parameters $Q_{X}^{s, s^\prime}$ and $Q^{s, s^\prime}_{\psi}$  decrease with the interaction and   vanish abruptly  in  the strong-coupling regime. In this regime, we also have $Z = 0$ and the   self-consistent equations simplify to
\begin{gather}
n_F(- \mu + \vert M \vert) + n_F(- \mu - \vert M \vert) = 1, \\
\frac{\sqrt{2} U}{\Delta_{X}} \coth \left(\frac{\beta \Delta_{X}}{2 \sqrt{2}} \right) = 1, \\
M = \frac{U}{2} \mathrm{sgn}(M)[n_F(- \mu - \vert M \vert) - n_F(- \mu + \vert M \vert)],
\end{gather}
where $n_F(z) \equiv (e^{\beta z} + 1)^{-1}$ is the Fermi-Dirac distribution function. The first equation yields  the chemical potential  $\mu = 0$, while at   low temperatures ($T \ll t_1$) the staggered magnetization and the Mott gap approach   $\vert M \vert = U/2$ and $\Delta_X = \sqrt{2} U$. As a result, the spinon dispersions $E^{\pm}_{\sigma, \psi}(\mathbf{k})$ become flat, leading to a constant spinon gap $\min_{\mathbf{k}}[E^{+}_{\sigma, \psi}(\mathbf{k}) - E^{-}_{\sigma, \psi}(\mathbf{k})] = U$, which coincides with  the chargon mass $m_X \equiv \Delta^2_X/(2 U)$, as predicted for standard theories of Mott insulators \cite{Imada-RMP(1998)}. The vanishing of the renormalized hopping amplitudes in this regime, labeled    $\mathrm{A\ell MI}^*$ in Fig. \ref{Fig2},  has also been observed in other models and may be an artifact of the saddle-point approximation \cite{Ko-PRB(2011)}. Nevertheless, as a general trend, the slave-rotor theory indicates that the $d$-wave spin splitting is substantially suppressed for $U \gg t_1$.    In this strong-coupling regime, altermagnetism can still manifest itself in the splitting of magnon bands, as described by effective spin models \cite{Smejkal_magnons(2023),Zhu-PRL2025,Consoli-PRL(2025)}.

\section{Electronic spectral features}\label{SecIV}
\noindent

To demonstrate that electronic excitations have altermagnetic properties, as shown for spinons, we now proceed to calculate the electron spectral function $\mathcal{A}_{\mathbf{k}}(\omega) = - \frac{1}{\pi} \operatorname{Im} \lbrace \operatorname{tr} [G(\mathbf{k}, \omega + i0^{+})]\rbrace$. The matrix elements of the retarded Green's function $G(\mathbf{k}, \omega + i0^{+})$ are obtained by Fourier transforming $G^{s s^{\prime}}_{\sigma}(\boldsymbol{r} - \boldsymbol{r}^{\prime}, \tau) = - \langle T_{\tau} \lbrace e^{i [\theta^s_{\boldsymbol{r}}(\tau) - \theta^{s^{\prime}}_{\boldsymbol{r}^{\prime}}(0)]} \psi^s_{\boldsymbol{r}, \sigma}(\tau) \psi^{s^{\prime} \dagger}_{\boldsymbol{r}^{\prime}, \sigma^{\prime}}(0) \rbrace \rangle$. In the saddle-point approximation, the momentum- and Matsubara-frequency-dependent Green's function $G^{s s^{\prime}}_{\sigma}(\mathbf{k}, i\omega_n)$ is given by the convolution formula \cite{Georges-PRB(2002),Georges-PRB(2004),He-PRB(2022),Sangiovanni-PRL(2024)}
\begin{align}\label{Eq_ElecProp}
G^{s s^{\prime}}_{\sigma}(\mathbf{k}, i\omega_n) = & \; Z G^{s s^{\prime}}_{\sigma, \psi}(\mathbf{k}, i\omega_n) + \frac{T}{N_c} \sum_{\mathbf{q} \neq \mathbf{0}} \sum_{\nu_m} G^{s s^{\prime}}_X(\mathbf{q}, i\nu_m) \nonumber \\
& \times G^{s s^{\prime}}_{\sigma, \psi}(\mathbf{k} - \mathbf{q}, i\omega_n - i\nu_m),
\end{align}
where $G_{X}(\mathbf{k}, i\omega_n)$ and $G_{\sigma, \psi}(\mathbf{k}, i\omega_n)$ denote the chargon and spinon propagators, respectively (see Appendix \ref{AppB}).

Our results for the spectral function $\mathcal{A}_{\mathbf{k}}(\omega)$ are shown in Fig. \ref{Fig4}. In the $\mathrm{A\ell M}$ phase, the electron dispersion extracted from the single-particle peaks in the spectral function matches the spinon dispersion, as can  be seen in the spin-split electron Fermi surface in Fig. \ref{Fig4}(a) and  in the energy dependence  in Fig. \ref{Fig4}(b). Similar results were obtained for fractionalized altermagnets in Ref. \cite{Sobral-PRR(2025)}. Using spin-resolved spectral functions $\mathcal{A}_{\mathbf{k}, \sigma}(\omega)$, we find that electron spin splitting decreases progressively as   $U$ increases [see Figs. \ref{Fig4}(c)--\ref{Fig4}(e)], as observed for spinon excitations. Remarkably, within the $\mathrm{A\ell MI}$ phase a small spin splitting can still be observed in the continuum. We also note that, beyond the saddle-point approximation employed here,  chargon-spinon interactions can give rise to a bound state associated with a spin polaron peak in the spectral function \cite{Daghofer-PRL(2026),Lanzini-PRL(2026)}. Finally, the behavior of the density of states $N(\omega) = \int_{\mathrm{BZ}}\frac{d^2\mathbf{k}}{(2\pi)^2} \mathcal{A}_{\mathbf{k}}(\omega)$ shown in Fig. \ref{Fig4}(f) is consistent with the metallic and insulating phases in this model.

\section{Summary and outlook}\label{SecV}
\noindent

We performed  a slave-rotor analysis of the ground-state phase diagram of a half-filled altermagnetic Hubbard model  on the Lieb lattice. At small repulsive interactions $U$, the model behaves as a normal metal. As $U$ increases, the system  undergoes a series of quantum phase transitions to altermagnetic states displaying metallic, insulating, and Mott-insulating behaviors. Due to the geometry of the Lieb lattice, all altermagnetic phases exhibit $d_{x^2 - y^2}$-wave spin splitting. While the magnetization order parameter increases with the interaction strength, the spin splitting observed in the electron spectral function decreases with $U$ and disappears altogether in the strong-coupling regime. We argued that this behavior can limit the observation of spin splitting in $d$-wave altermagnetic candidates with an underlying Lieb lattice to systems exhibiting weak-to-moderate electronic correlations. At this point, we mention that these features rely on the distinct effect played by the spin and charge degrees of freedom governed by the Hubbard interaction, which are described, respectively, by the spinon and chargon sectors of the slave-rotor theory. The former are responsible for the appearance of the altermagnetic order, signaled by a finite staggered magnetization, while the latter promote the Mott-insulating state. Moreover, although all the self-consistent equations for the order parameters obtained here were derived within the saddle-point approximation, the independent field theories describing the staggered magnetization and the bond variables are in principle more appropriate for the weak- and strong-coupling regimes, respectively. In this regard, we point out that such a combined approach was performed, e.g., in Ref. \cite{Ko-PRB(2011)}, and is expected to capture the qualitative behavior of the system for a wide range of intermediate $U$, where the altermagnetic phase transitions uncovered in this work take place. This is also corroborated by a recent quantum Monte Carlo study \cite{Ma-arXiv(2026)}, which reported a similar suppression of the momentum-space spin splitting with increasing $U$ in a related altermagnetic Hubbard model.

Finally, we must add that, beyond the application to the oxychalcogenides, the quantum phase transitions discussed here could also be experimentally tested using ultracold atoms in optical Lieb-lattice setups \cite{Knap-PRL(2024),Greiner-S(2026)}. Extensions of this work might also include a systematic investigation of the phase diagram of the altermagnetic Hubbard model as a function of temperature and doping, as the slave-rotor approach may reveal unconventional fractionalized superconducting phases and pair-density-wave states \cite{Mazin-arXiv(2022),Ouassou-PRL(2023),Linder-PRB(2023),Neupert-NC(2024),Zhu-PRB(2023),Sudbo-PRB(2023),Chakraborty-PRB(2024),Carvalho-PRB(2024),Scheurer-PRB(2024),Schaffer-PRB(2025),Hong-PRB(2025),Knolle-PRB(2025),Scheurer-PRL(2025),Classen-PRB(2025),Wu-PRL(2025),Froldi-AP(2025)}.

\begin{acknowledgments}
We thank R. M. Fernandes for fruitful discussions that motivated this work. V.S.d.C. acknowledges funding from the  National Council for Scientific and Technological Development (CNPq) under Grants No. 404274/2023-4, No. 407658/2025-4, and No. 312661/2025-8. He also thanks funding from the Simons Foundation through the project ``IIP-Theoretical Research in Frontier Areas'' (Grant No. 1023171). H.F. acknowledges funding from the CNPq under Grants  No. 305575/2025-2, No. 404274/2023-4, No. 407658/2025-4, and from the Centre National de la Recherche Scientifique (CNRS) International Research Project ``Non-perturbative methods in strongly coupled field theories and statistics.'' R.G.P. acknowledges funding  from the Simons Foundation (Grant No. 1023171),   Finep (Grant No. 1699/24 IIFFINEP), and CNPq (Grants No. 309569/2022-2 and No. 404274/2023-4). Two of us (V.S.d.C. and H.F.) also acknowledge the support of the INCT project Advanced Quantum Materials, involving the Brazilian agencies CNPq (Grant No. 408766/2024-7), FAPESP (Grant No. 2025/27091-3), and CAPES.
\end{acknowledgments}

\section*{Data availability}
\noindent

There are no publicly available research data or software supporting this manuscript. Requests for further information or data should be sent to the authors.

\noindent 

\appendix

\onecolumngrid

\section{\uppercase{Slave-rotor theory applied to the altermagnetic Hubbard model}}\label{AppA}
\noindent

This appendix deals with the derivation of the self-consistent equations arising from the application of the slave-rotor theory to the altermagnetic Hubbard model defined in the main text. To do that, we start from the Lagrangian shown in Eq. \ref{SRotorLag}. After decoupling it in terms of the renormalized chargon and spinon hoppings $Q^{s s'}_{X}(\tau)$ and $Q^{s s'}_{\psi}(\tau)$ and the staggered magnetization $M(\mathbf{r}, \tau) = (- 1)^s M_{\boldsymbol{r}}(\tau)$ with the convention $(- 1)^s = + 1$ $(- 1)$ for $s \in A$ $(B)$, one obtains the effective Lagragian: 
\begin{align}
\mathcal{L}[\bar{\psi}, \psi, \bar{X}, X, h, \lambda, Q, M, \tau] = & \sum_{\boldsymbol{r}, s} \bigg\{ \bar{\psi}^{s}_{\boldsymbol{r}} (\partial_{\tau} - \mu - h^{s}_{\boldsymbol{r}})\psi^{s}_{\boldsymbol{r}} - \frac{(h^{s}_{\boldsymbol{r}})^2}{U} + h^{s}_{\boldsymbol{r}} + \frac{\vert \partial_{\tau} X^s_{\boldsymbol{r}}\vert^2}{U} + \frac{h^s_{\boldsymbol{r}}}{U} (\bar{X}^s_{\boldsymbol{r}} \partial_{\tau} X^s_{\boldsymbol{r}} - X^s_{\boldsymbol{r}} \partial_{\tau} \bar{X}^s_{\boldsymbol{r}}) \nonumber \\
& + \lambda_{\boldsymbol{r}}(\vert X^s_{\boldsymbol{r}} \vert^2 - 1) \bigg\rbrace + \mathcal{L}_{\mathcal{Q}}[\bar{\psi}, \psi, \bar{X}, X, Q, \tau] + \mathcal{L}_{\mathcal{M}}[\bar{\psi}, \psi, M, \tau], \label{Eq_Lag01}
\end{align}
where $\psi^{s}_{\boldsymbol{r}} \equiv (\psi^{s}_{\boldsymbol{r}, \uparrow}, \; \psi^{s}_{\boldsymbol{r}, \downarrow})^T$, and the last two terms on the right-hand side of Eq. \eqref{Eq_Lag01} refer to
\begin{align}
\mathcal{L}_{\mathcal{Q}}[\bar{\psi}, \psi, \bar{X}, X, Q, \tau] = & - t_1 \sum_{\boldsymbol{r}} \sum^4_{j = 1} [Q^{A B}_{\psi}(\tau) \bar{\psi}^A_{\boldsymbol{r}}(\tau) \psi^B_{\boldsymbol{r} + \boldsymbol{u}_j}(\tau) + Q^{A B}_{X}(\tau) \bar{X}^A_{\boldsymbol{r}}(\tau) X^B_{\boldsymbol{r} + \boldsymbol{u}_j}(\tau) + \mathrm{H. c.} ] \nonumber \\
& - t_{2a} \sum_{\boldsymbol{r}} [Q^{A A}_{\psi, 2a}(\tau) \bar{\psi}^A_{\boldsymbol{r}}(\tau) \psi^A_{\boldsymbol{r} + \boldsymbol{v}_2}(\tau) + Q^{B B}_{\psi, 2a}(\tau) \bar{\psi}^B_{\boldsymbol{r}}(\tau) \psi^{B}_{\boldsymbol{r} + \boldsymbol{v}_1}(\tau) \nonumber \\
& + Q^{A A}_{X, 2a}(\tau) \bar{X}^A_{\boldsymbol{r}}(\tau) X^A_{\boldsymbol{r} + \boldsymbol{v}_2}(\tau) + Q^{B B}_{X, 2a}(\tau) \bar{X}^B_{\boldsymbol{r}}(\tau) X^{B}_{\boldsymbol{r} + \boldsymbol{v}_1}(\tau) + \mathrm{H. c.}] \nonumber \\
& - t_{2b} \sum_{\boldsymbol{r}} [Q^{A A}_{\psi, 2b}(\tau) \bar{\psi}^A_{\boldsymbol{r}}(\tau) \psi^A_{\boldsymbol{r} + \boldsymbol{v}_1}(\tau) + Q^{B B}_{\psi, 2b}(\tau) \bar{\psi}^B_{\boldsymbol{r}}(\tau) \psi^{B}_{\boldsymbol{r} + \boldsymbol{v}_2}(\tau) \nonumber \\
& + Q^{A A}_{X, 2b}(\tau) \bar{X}^A_{\boldsymbol{r}}(\tau) X^A_{\boldsymbol{r} + \boldsymbol{v}_1}(\tau) + Q^{B B}_{X, 2b}(\tau) \bar{X}^B_{\boldsymbol{r}}(\tau) X^{B}_{\boldsymbol{r} + \boldsymbol{v}_2}(\tau) + \mathrm{H. c.}] \nonumber \\
& + 4 t_1 N_{c} [Q^{A B}_{\psi}(\tau) Q^{A B}_{X}(\tau) + Q^{B A}_{\psi}(\tau) Q^{B A}_{X}(\tau)] \nonumber \\
& + 2 t_{2a} N_{c} [Q^{A A}_{\psi, 2a}(\tau) Q^{A A}_{X, 2a}(\tau) + Q^{B B}_{\psi, 2a}(\tau) Q^{B B}_{X, 2a}(\tau)] \nonumber \\
& + 2 t_{2b} N_{c} [Q^{A A}_{\psi, 2b}(\tau) Q^{A A}_{X, 2b}(\tau) + Q^{B B}_{\psi, 2b}(\tau) Q^{B B}_{X, 2b}(\tau)], \label{Eq_Lag02}\\
\mathcal{L}_{\mathcal{M}}[\bar{\psi}, \psi, M, \tau] = & - \sum_{\boldsymbol{r}, s} (- 1)^s M_{\boldsymbol{r}}(\tau) [\bar{\psi}^s_{\boldsymbol{r}}(\tau) \sigma^z \psi^s_{\boldsymbol{r}}(\tau)] + \sum_{\boldsymbol{r}, s} \frac{M^2_{\boldsymbol{r}}(\tau)}{U}. \label{Eq_Lag03}
\end{align}
The lattice vectors $\boldsymbol{u}_j$ ($\boldsymbol{v}_j$) that appear in the latter equations connect NN (NNN) sites of a Lieb lattice with $N_{c}$ unit cells. According to Fig. 1 in the main text, these vectors are defined by
\begin{equation}
    \boldsymbol{u}_1 = \frac{a}{2}(\hat{\boldsymbol{x}} + \hat{\boldsymbol{y}}), \hspace{0.5cm} \boldsymbol{u}_2 = \frac{a}{2}( - \hat{\boldsymbol{x}} + \hat{\boldsymbol{y}}), \hspace{0.5cm} \boldsymbol{u}_3 = \frac{a}{2}( - \hat{\boldsymbol{x}} - \hat{\boldsymbol{y}}), \hspace{0.5cm} \boldsymbol{u}_4 = \frac{a}{2}(\hat{\boldsymbol{x}} - \hat{\boldsymbol{y}}), \hspace{0.5cm}  \boldsymbol{v}_1 = a \hat{\boldsymbol{x}}, \hspace{0.5cm} \boldsymbol{v}_2 = a \hat{\boldsymbol{y}}.
\end{equation}

In terms of the orders parameters introduced so far, the partition function of the altermagnetic Hubbard model assumes the form
\begin{equation}
\mathcal{Z} = \int \mathscr{D}[h, \lambda, Q, M] e^{- \mathcal{S}[h, \lambda, Q, M]}, 
\end{equation}
where the equation
\begin{equation}
e^{- \mathcal{S}[h, \lambda, Q, M]} = \int \mathscr{D}[\bar{\psi}, \psi, \bar{X}, X] \exp\bigg\lbrace - \int^{\beta}_{0} d\tau \mathcal{L}[\bar{\psi}, \psi, \bar{X}, X, h, \lambda, Q, M, \tau] \bigg\rbrace
\end{equation}
defines the action $\mathcal{S}[h, \lambda, Q, M]$ whose minimization   determines the self-consistent (or saddle-point) equations obeyed by the order parameters and Lagrange multipliers. Indeed, the minimization with respect to the Lagrange multipliers,
\begin{equation}
\frac{\delta \mathcal{S}[h, \lambda, Q, M]}{\delta h^{s}_{\boldsymbol{r}}(\tau)}\bigg\vert_{h^{s}_{\boldsymbol{r}}(\tau) = h^{s}} = 0, \hspace{0.5cm} \frac{\delta \mathcal{S}[h, \lambda, Q, M]}{\delta \lambda_{\boldsymbol{r}}(\tau)}\bigg\vert_{\lambda_{\boldsymbol{r}}(\tau) = \lambda} = 0,
\end{equation}
leads to
\begin{align}
\frac{T}{N_{c}} \sum_{\sigma = \uparrow, \downarrow} \sum_{\mathbf{k}} \sum_{\omega_n} G^{ss}_{\sigma, \psi}(\mathbf{k}, i\omega_n) e^{i\omega_n 0^{+}} & = 1 - \frac{2 h^s}{U} - \frac{T}{U N_{c}} \sum_{\mathbf{k} \neq \mathbf{0}} \sum_{\nu_n} i\nu_n G^{ss}_X(\mathbf{k}, i\nu_n)(e^{i\nu_n 0^{+}} + e^{- i\nu_n 0^{+}}), \label{Eq_LagMult01}\\
\frac{T}{2 N_{c}}\sum_{\mathbf{k} \neq \mathbf{0}} \sum_{s = A, B} \sum_{\nu_n} G^{ss}_{X}(\mathbf{k}, i\nu_n)e^{i\nu_n 0^{+}} & = 1 - Z. \label{Eq_LagMult02}
\end{align}
Here, $G^{ss}_X(\mathbf{k}, i\nu_n)$ and $G^{ss'}_{\sigma, \psi}(\mathbf{k}, i\omega_n)$  correspond, respectively, to the chargon and spin Green's functions in momentum-frequency space, which are   obtained by Fourier transforming the propagators $G^{ss'}_X(\boldsymbol{r} - \boldsymbol{r}', \tau) = \langle T_{\tau}[X^s_{\boldsymbol{r}, \sigma}(\tau) \bar{X}^{s'}_{\boldsymbol{r}', \sigma}(0)]\rangle$ and $G^{ss'}_{\sigma, \psi}(\boldsymbol{r} - \boldsymbol{r}', \tau) = - \langle T_{\tau}[\psi^s_{\boldsymbol{r}, \sigma}(\tau) \bar{\psi}^{s'}_{\boldsymbol{r}', \sigma}(0)]\rangle$. In addition, $Z$ denotes the quasiparticle weight (or amplitude of the Bose condensation), which is finite in the Fermi-liquid regime   but vanishes at the Mott transition \cite{Georges-PRB(2002),Georges-PRB(2004),Imada-RMP(1998)}. Since the Hubbard interaction is the same for electrons located in different   sublattices, we assumed in the derivation of Eq. \eqref{Eq_LagMult02} that $Z$ is uniform, which   implies $Z = Z^s$ where $\sqrt{Z^s Z^{s'}}\equiv \langle T_{\tau}[X^s_{\mathbf{k} = \mathbf{0}}(\tau) X^{s' \dagger}_{\mathbf{k} = \mathbf{0}}(\tau)]\rangle/N_{c}$ \cite{Lee-PRL(2005),Sangiovanni-PRL(2024)}.

The remaining slave-rotor equations are found by imposing the conditions
\begin{align}
\frac{\delta \mathcal{S}[h, \lambda, Q, M]}{\delta Q^{s s'}_{\psi/X}(\tau)}\bigg\vert_{Q^{s s'}_{\psi/X}(\tau) = Q^{s s'}_{\psi/X}} & = 0, \hspace{0.5cm} \frac{\delta \mathcal{S}[h, \lambda, Q, M]}{\delta M_{\boldsymbol{r}}(\tau)}\bigg\vert_{M_{\boldsymbol{r}}(\tau) = M} = 0, \\
\frac{\delta \mathcal{S}[h, \lambda, Q, M]}{\delta Q^{s s}_{\psi/X, 2a}(\tau)}\bigg\vert_{Q^{s s}_{\psi/X, 2a}(\tau) = Q^{s s}_{\psi/X, 2a}} & = 0, \hspace{0.5cm} \frac{\delta \mathcal{S}[h, \lambda, Q, M]}{\delta Q^{s s}_{\psi/X, 2b}(\tau)}\bigg\vert_{Q^{s s}_{\psi/X, 2b}(\tau) = Q^{s s}_{\psi/X, 2b}} = 0.
\end{align}
It is straightforward to show that they yield
\begin{gather}
Q^{A B}_X = \frac{T}{4 N_{c}} \sum_{\sigma = \uparrow, \downarrow} \sum_{\mathbf{k}} \sum_{\omega_n} V_{\mathbf{k}} G^{B A}_{\sigma, \psi}(\mathbf{k}, i\omega_n) e^{i\omega_n 0^{+}}, \label{Eq_OP_First}\\
Q^{A B}_{\psi} = Z + \frac{T}{4 N_{c}} \sum_{\mathbf{k} \neq \mathbf{0}} \sum_{\nu_n} V_{\mathbf{k}} G^{B A}_X(\mathbf{k}, i\nu_n) e^{i\nu_n 0^{+}}, \\
Q^{A A}_{X, 2a} = \frac{T}{N_{c}} \sum_{\sigma = \uparrow, \downarrow} \sum_{\mathbf{k}} \sum_{\omega_n} \cos(k_y a) G^{A A}_{\sigma, \psi}(\mathbf{k}, i\omega_n) e^{i\omega_n 0^{+}}, \\
Q^{A A}_{\psi, 2a} = Z + \frac{T}{N_{c}} \sum_{\mathbf{k} \neq \mathbf{0}} \sum_{\nu_n} \cos(k_y a) G^{A A}_X(\mathbf{k}, i\nu_n) e^{i\nu_n 0^{+}}, \\
Q^{B B}_{X, 2a} = \frac{T}{N_{c}} \sum_{\sigma = \uparrow, \downarrow} \sum_{\mathbf{k}} \sum_{\omega_n} \cos(k_x a) G^{B B}_{\sigma, \psi}(\mathbf{k}, i\omega_n) e^{i\omega_n 0^{+}},\\
Q^{B B}_{\psi, 2a} = Z + \frac{T}{N_{c}} \sum_{\mathbf{k} \neq \mathbf{0}} \sum_{\nu_n} \cos(k_x a) G^{B B}_X(\mathbf{k}, i\nu_n) e^{i\nu_n 0^{+}}, \\
Q^{A A}_{X, 2b} = \frac{T}{N_{c}} \sum_{\sigma = \uparrow, \downarrow} \sum_{\mathbf{k}} \sum_{\omega_n} \cos(k_x a) G^{A A}_{\sigma, \psi}(\mathbf{k}, i\omega_n) e^{i\omega_n 0^{+}}, \\
Q^{A A}_{\psi, 2b} = Z + \frac{T}{N_{c}} \sum_{\mathbf{k} \neq \mathbf{0}} \sum_{\nu_n} \cos(k_x a) G^{A A}_X(\mathbf{k}, i\nu_n) e^{i\nu_n 0^{+}}, \\
Q^{B B}_{X, 2b} = \frac{T}{N_{c}} \sum_{\sigma = \uparrow, \downarrow} \sum_{\mathbf{k}} \sum_{\omega_n} \cos(k_y a) G^{B B}_{\sigma, \psi}(\mathbf{k}, i\omega_n) e^{i\omega_n 0^{+}}, \\
Q^{B B}_{\psi, 2b} = Z + \frac{T}{N_{c}} \sum_{\mathbf{k} \neq \mathbf{0}} \sum_{\nu_n} \cos(k_y a) G^{B B}_X(\mathbf{k}, i\nu_n) e^{i\nu_n 0^{+}}, \\
M = \frac{U T}{4 N_{c}} \sum_{\sigma = \uparrow, \downarrow}\sum_{\mathbf{k}} \sum_{s = A, B} \sum_{\omega_n} (-1)^{s + \sigma} G^{ss}_{\sigma, \psi}(\mathbf{k}, i\omega_n) e^{i\omega_n 0^{+}}, \label{Eq_OP_Last}
\end{gather}
where $Q^{B A}_X = Q^{A B *}_X$, $Q^{B A}_{\psi} = Q^{A B *}_{\psi}$, and $V_{\mathbf{k}} \equiv \sum^{4}_{j = 1} e^{i \mathbf{k} \cdot \boldsymbol{u}_j}$. To obtain the self-consistent equation for the staggered magnetization $M$, we adopted the convention $(- 1)^{\sigma} = + 1$ $(-1)$ for $\sigma = \; \uparrow$ $(\downarrow)$.

To solve the self-consistent equations, we first determine the analytical expressions for the Green's functions $G^{ss}_X(\mathbf{k}, i\nu_n)$ and $G^{ss'}_{\sigma, \psi}(\mathbf{k}, i\omega_n)$. In the saddle-point approximation, Eqs. \eqref{Eq_Lag01}--\eqref{Eq_Lag03} yield
\begin{align}
G^{- 1}_{\sigma, \psi}(\mathbf{k}, i\omega_n) & =
\begin{pmatrix}
i \omega_n - [\epsilon^{A A}_{\psi}(\mathbf{k}) - \mu] + (- 1)^{\sigma} M + h^A & t_1 Q^{A B}_{\psi} V_{\mathbf{k}} \\
t_1 Q^{B A}_{\psi} V^{*}_{\mathbf{k}} & i \omega_n - [\epsilon^{B B}_{\psi}(\mathbf{k}) - \mu] - (- 1)^{\sigma} M + h^B
\end{pmatrix}, \\
G^{- 1}_X(\mathbf{k}, i\nu_n) & =
\begin{pmatrix}
\dfrac{\nu^2_n}{U} - \dfrac{2 i \nu_n h^A}{U} + \epsilon^{A A}_X(\mathbf{k}) + \lambda & - t_1 Q^{A B}_X V_{\mathbf{k}} \\
- t_1 Q^{B A}_X V^{*}_{\mathbf{k}} & \dfrac{\nu^2_n}{U} - \dfrac{2 i \nu_n h^B}{U} + \epsilon^{B B}_X(\mathbf{k}) + \lambda
\end{pmatrix},
\end{align}
which depend on the dispersions
\begin{align}
\epsilon^{A A}_{\psi}(\mathbf{k}) & = - 2 t_{2a} Q^{A A}_{\psi, 2a} \cos(k_y a) - 2 t_{2b} Q^{A A}_{\psi, 2b} \cos(k_x a), \\
\epsilon^{B B}_{\psi}(\mathbf{k}) & = - 2 t_{2a} Q^{B B}_{\psi, 2a} \cos(k_x a) - 2 t_{2b} Q^{B B}_{\psi, 2b} \cos(k_y a), \\
\epsilon^{A A}_X(\mathbf{k}) & = - 2 t_{2a} Q^{A A}_{X, 2a} \cos(k_y a) - 2 t_{2b} Q^{A A}_{X, 2b} \cos(k_x a), \\
\epsilon^{B B}_X(\mathbf{k}) & = - 2 t_{2a} Q^{B B}_{X, 2a} \cos(k_x a) - 2 t_{2b} Q^{B B}_{X, 2b} \cos(k_y a).
\end{align}
Clearly, the order parameters $Q^{s s^{\prime}}_{X}$ and $Q^{s s^{\prime}}_{\psi}$ produce the renormalization of the NN and NNN hoppings for the chargons and spinons, respectively.

Hereafter, we focus on half-filling. In this case, we immediately find $h^A = h^B = 0$. In addition, after evaluating the Matsubara sums in Eqs. \eqref{Eq_LagMult01} and \eqref{Eq_LagMult02}, we obtain
\begin{gather}
\frac{1}{N_{c}} \sum_{\sigma = \uparrow, \downarrow} \sum_{\mathbf{k}} \lbrace n_F[E^{+}_{\sigma, \psi}(\mathbf{k})] + n_F[E^{-}_{\sigma, \psi}(\mathbf{k})] \rbrace = 2, \\
\frac{U}{4 N_{c}} \sum_{\mathbf{k} \neq \mathbf{0}} \bigg\lbrace \frac{1}{E^{+}_X(\mathbf{k})} \coth\bigg[ \frac{\beta E^{+}_X(\mathbf{k})}{2} \bigg] + \frac{1}{E^{-}_X(\mathbf{k})} \coth\bigg[ \frac{\beta E^{-}_X(\mathbf{k})}{2} \bigg] \bigg\rbrace = 1 - Z,
\end{gather}
where $n_F(z) = (e^{\beta z} + 1)^{-1}$ is the Fermi-Dirac distribution function, and $E^{\pm}_{\sigma, \psi}(\mathbf{k})$ denotes the spinon dispersions. They correspond to the poles of $G_{\sigma, \psi}(\mathbf{k}, i\omega_n)$ and can be expressed as
\begin{gather}
E^{\pm}_{\sigma, \psi}(\mathbf{k}) = \frac{1}{2}\left\lbrace E^{A A}_{\sigma, \psi}(\mathbf{k}) + E^{B B}_{\sigma, \psi}(\mathbf{k}) \pm \sqrt{[E^{A A}_{\sigma, \psi}(\mathbf{k}) - E^{B B}_{\sigma, \psi}(\mathbf{k})]^2 + (2 t_1 \vert Q^{A B}_{\psi} V_{\mathbf{k}} \vert)^2} \right\rbrace, \label{Eq_SpinonD01}\\
E^{A A}_{\sigma, \psi}(\mathbf{k}) = \epsilon^{A A}_{\sigma, \psi}(\mathbf{k}) - \mu - (- 1)^{\sigma} M, \label{Eq_SpinonD02} \\
E^{B B}_{\sigma, \psi}(\mathbf{k}) = \epsilon^{B B}_{\sigma, \psi}(\mathbf{k}) - \mu + (- 1)^{\sigma} M. \label{Eq_SpinonD03}
\end{gather}
Likewise, $G_{X}(\mathbf{k}, i\omega_n)$ has poles at the chargon dispersions $E^{\pm}_X(\mathbf{k})$, which are given by
\begin{align}
E^{\pm}_X(\mathbf{k}) = \sqrt{\frac{U}{2}} \sqrt{E^{A A}_X(\mathbf{k}) + E^{B B}_X(\mathbf{k}) \pm \sqrt{[E^{A A}_X(\mathbf{k}) - E^{B B}_X(\mathbf{k})]^2 + (2 t_1 \vert Q^{A B}_X V_{\mathbf{k}} \vert)^2}},
\end{align}
with $E^{s s}(\mathbf{k}) = \epsilon^{s s}(\mathbf{k}) + \lambda$. Note that $E^{\pm}_X(\mathbf{k})$ must be real-valued functions since the bosons describing the chargon excitations are assumed to have quantum-mechanical stability \cite{Georges-PRB(2002),Georges-PRB(2004),Huang-PRB(2020)}. This condition is satisfied by a careful definition of the Lagrange multiplier $\lambda$. We find
\begin{equation}
\lambda = - \frac{1}{2} \min_{\mathbf{k} \in \mathrm{BZ}}\left\lbrace \epsilon^{A A}_X(\mathbf{k}) + \epsilon^{B B}_X(\mathbf{k}) - \sqrt{[\epsilon^{A A}_X(\mathbf{k}) - \epsilon^{B B}_X(\mathbf{k})]^2 + (2 t_1 \vert Q^{A B}_X V_{\mathbf{k}} \vert)^2} \right\rbrace + \frac{\Delta^2_X}{4 U},
\end{equation}
where the momentum $\mathbf{k}$ runs over the Brillouin zone (BZ) and $\Delta_X \in \mathbb{R}_{+}$ refers to the Mott gap. Moreover, one finds  $\Delta_X \neq 0$  in the Mott-insulating phase, where the quasiparticle weight $Z$ vanishes.

\begin{figure}[t]
\centering
\includegraphics[width=0.85\linewidth,valign=t]{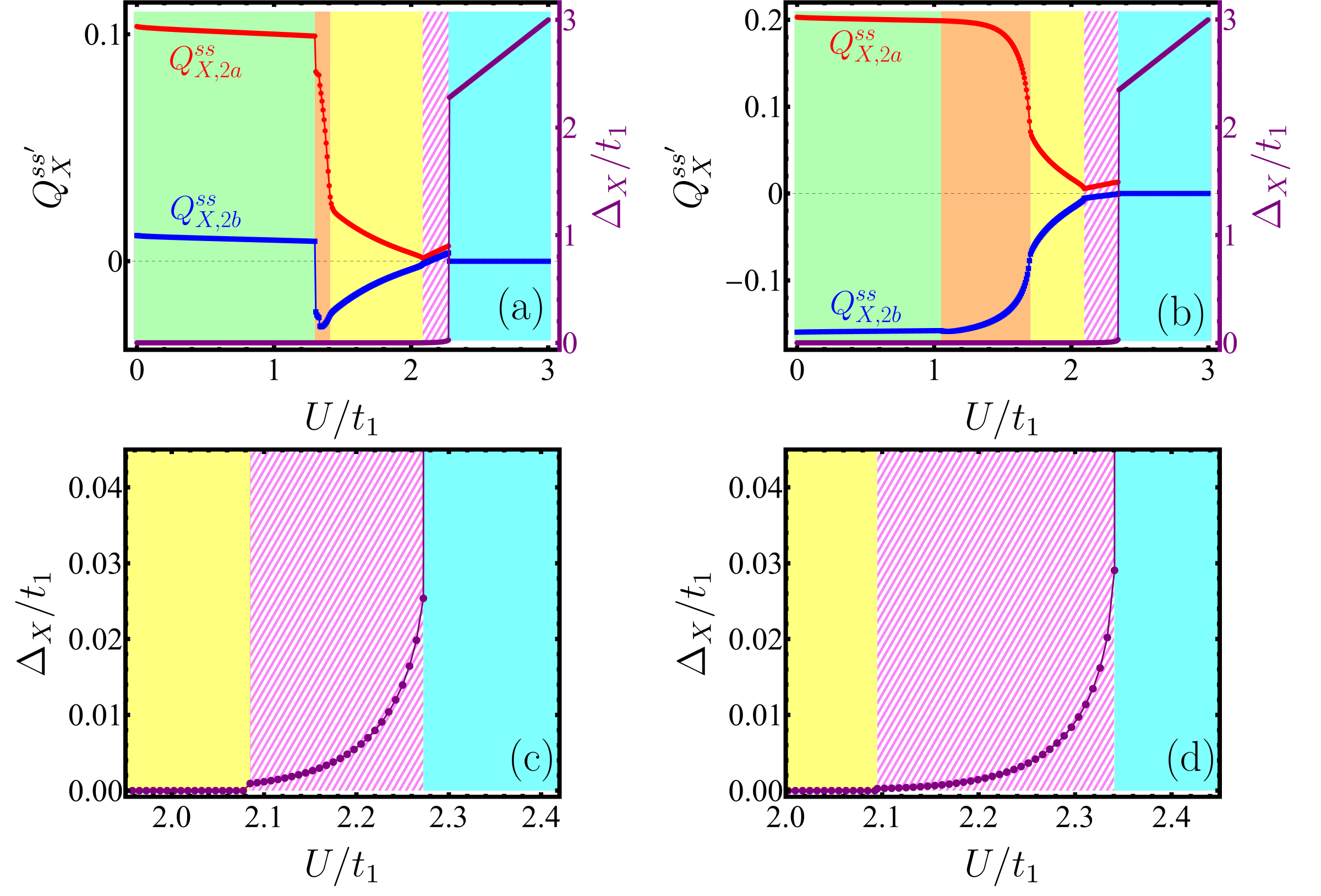}
\caption{Behavior of the renormalized chargon NNN hoppings and the Mott gap $\Delta_X$ as a function of the Hubbard interaction $U$. The dependence of the bare NNN hoppings is parametrized by $t_{2a/2b} = t^{\prime}(1 \pm \delta)$, such that we fix $t^{\prime} = 0.3 t_1$ and vary the anisotropic parameter as (a)--(c) $\delta = 0.2$ and (b)--(d) $\delta = 0.8$. The color scheme of the different phases of the model follows that employed in the main text. Notice that $\Delta_X$ starts to differ from zero after the system reaches the $\mathrm{A\ell MI}$ state (magenta hatched region).}\label{Fig5}
\end{figure}

Using these  results, we can finally evaluate the Matsubara sums defining the self-consistent equations \eqref{Eq_OP_First}--\eqref{Eq_OP_Last}. We obtain
\begin{align}
Q^{A B}_X = & - \frac{t_1 Q^{B A}_{\psi}}{4 N_{c}} \sum_{\sigma = \uparrow, \downarrow} \sum_{\mathbf{k}} \vert V_{\mathbf{k}} \vert^2 \bigg\lbrace \frac{n_F[E^{+}_{\sigma, \psi}(\mathbf{k})] - n_F[E^{-}_{\sigma, \psi}(\mathbf{k})]}{E^{+}_{\sigma, \psi}(\mathbf{k}) - E^{-}_{\sigma, \psi}(\mathbf{k})} \bigg\rbrace, \\
Q^{A B}_{\psi} = & \; Z -  \frac{t_1 U^2 Q^{B A}_X}{8 N_{c}} \sum_{\mathbf{k} \neq \mathbf{0}} \frac{\vert V_{\mathbf{k}} \vert^2}{[E^{+}_X(\mathbf{k})]^2 - [E^{-}_X(\mathbf{k})]^2} \bigg\lbrace \frac{1}{E^{+}_X(\mathbf{k})} \coth\bigg[ \frac{\beta E^{+}_X(\mathbf{k})}{2} \bigg] + \frac{1}{E^{-}_X(\mathbf{k})} \coth\bigg[ \frac{\beta E^{-}_X(\mathbf{k})}{2} \bigg] \bigg\rbrace, \\
Q^{A A}_{X, 2a} = & \; \frac{1}{N_{c}} \sum_{\sigma = \uparrow, \downarrow} \sum_{\mathbf{k}} \cos(k_y a) \bigg\lbrace \bigg[ \frac{E^{+}_{\sigma, \psi}(\mathbf{k}) - E^{B B}_{\sigma, \psi}(\mathbf{k})}{E^{+}_{\sigma, \psi}(\mathbf{k}) - E^{-}_{\sigma, \psi}(\mathbf{k})} \bigg] n_F[E^{+}_{\sigma, \psi}(\mathbf{k})] + \bigg[ \frac{E^{-}_{\sigma, \psi}(\mathbf{k}) - E^{B B}_{\sigma, \psi}(\mathbf{k})}{E^{-}_{\sigma, \psi}(\mathbf{k}) - E^{+}_{\sigma, \psi}(\mathbf{k})} \bigg] n_F[E^{-}_{\sigma, \psi}(\mathbf{k})] \bigg\rbrace, \\
Q^{A A}_{\psi, 2a} = & \; Z + \frac{U}{2 N_{c}} \sum_{\mathbf{k} \neq \mathbf{0}} \frac{\cos(k_y a)}{[E^{+}_X(\mathbf{k})]^2 - [E^{-}_X(\mathbf{k})]^2} \bigg\lbrace \bigg[ \frac{[E^{+}_X(\mathbf{k})]^2 - U E^{B B}_X(\mathbf{k})}{E^{+}_X(\mathbf{k})} \bigg] \coth\bigg[ \frac{\beta E^{+}_X(\mathbf{k})}{2} \bigg] + \bigg[ \frac{[E^{-}_X(\mathbf{k})]^2 - U E^{B B}_X(\mathbf{k})}{E^{-}_X(\mathbf{k})} \bigg] \nonumber \\
& \times \coth\bigg[ \frac{\beta E^{-}_X(\mathbf{k})}{2} \bigg] \bigg\rbrace, \\
Q^{B B}_{X, 2a} = & \; \frac{1}{N_{c}} \sum_{\sigma = \uparrow, \downarrow} \sum_{\mathbf{k}} \cos(k_x a) \bigg\lbrace \bigg[ \frac{E^{+}_{\sigma, \psi}(\mathbf{k}) - E^{A A}_{\sigma, \psi}(\mathbf{k})}{E^{+}_{\sigma, \psi}(\mathbf{k}) - E^{-}_{\sigma, \psi}(\mathbf{k})} \bigg] n_F[E^{+}_{\sigma, \psi}(\mathbf{k})] + \bigg[ \frac{E^{-}_{\sigma, \psi}(\mathbf{k}) - E^{A A}_{\sigma, \psi}(\mathbf{k})}{E^{-}_{\sigma, \psi}(\mathbf{k}) - E^{+}_{\sigma, \psi}(\mathbf{k})} \bigg] n_F[E^{-}_{\sigma, \psi}(\mathbf{k})] \bigg\rbrace, \\
Q^{B B}_{\psi, 2a} = & \; Z + \frac{U}{2 N_{c}} \sum_{\mathbf{k} \neq \mathbf{0}} \frac{\cos(k_x a)}{[E^{+}_X(\mathbf{k})]^2 - [E^{-}_X(\mathbf{k})]^2} \bigg\lbrace \bigg[ \frac{[E^{+}_X(\mathbf{k})]^2 - U E^{A A}_X(\mathbf{k})}{E^{+}_X(\mathbf{k})} \bigg] \coth\bigg[ \frac{\beta E^{+}_X(\mathbf{k})}{2} \bigg] + \bigg[ \frac{[E^{-}_X(\mathbf{k})]^2 - U E^{A A}_X(\mathbf{k})}{E^{-}_X(\mathbf{k})} \bigg] \nonumber \\
& \times \coth\bigg[ \frac{\beta E^{-}_X(\mathbf{k})}{2} \bigg] \bigg\rbrace, 
\end{align}

\begin{align}
Q^{A A}_{X, 2b} = & \; \frac{1}{N_{c}} \sum_{\sigma = \uparrow, \downarrow} \sum_{\mathbf{k}} \cos(k_x a) \bigg\lbrace \bigg[ \frac{E^{+}_{\sigma, \psi}(\mathbf{k}) - E^{B B}_{\sigma, \psi}(\mathbf{k})}{E^{+}_{\sigma, \psi}(\mathbf{k}) - E^{-}_{\sigma, \psi}(\mathbf{k})} \bigg] n_F[E^{+}_{\sigma, \psi}(\mathbf{k})] + \bigg[ \frac{E^{-}_{\sigma, \psi}(\mathbf{k}) - E^{B B}_{\sigma, \psi}(\mathbf{k})}{E^{-}_{\sigma, \psi}(\mathbf{k}) - E^{+}_{\sigma, \psi}(\mathbf{k})} \bigg] n_F[E^{-}_{\sigma, \psi}(\mathbf{k})] \bigg\rbrace, \\
Q^{A A}_{\psi, 2b} = & \; Z + \frac{U}{2 N_{c}} \sum_{\mathbf{k} \neq \mathbf{0}} \frac{\cos(k_x a)}{[E^{+}_X(\mathbf{k})]^2 - [E^{-}_X(\mathbf{k})]^2} \bigg\lbrace \bigg[ \frac{[E^{+}_X(\mathbf{k})]^2 - U E^{B B}_X(\mathbf{k})}{E^{+}_X(\mathbf{k})} \bigg] \coth\bigg[ \frac{\beta E^{+}_X(\mathbf{k})}{2} \bigg] + \bigg[ \frac{[E^{-}_X(\mathbf{k})]^2 - U E^{B B}_X(\mathbf{k})}{E^{-}_X(\mathbf{k})} \bigg] \nonumber \\
& \times \coth\bigg[ \frac{\beta E^{-}_X(\mathbf{k})}{2} \bigg] \bigg\rbrace, \\
Q^{B B}_{X, 2b} = & \; \frac{1}{N_{c}} \sum_{\sigma = \uparrow, \downarrow} \sum_{\mathbf{k}} \cos(k_y a) \bigg\lbrace \bigg[ \frac{E^{+}_{\sigma, \psi}(\mathbf{k}) - E^{A A}_{\sigma, \psi}(\mathbf{k})}{E^{+}_{\sigma, \psi}(\mathbf{k}) - E^{-}_{\sigma, \psi}(\mathbf{k})} \bigg] n_F[E^{+}_{\sigma, \psi}(\mathbf{k})] + \bigg[ \frac{E^{-}_{\sigma, \psi}(\mathbf{k}) - E^{A A}_{\sigma, \psi}(\mathbf{k})}{E^{-}_{\sigma, \psi}(\mathbf{k}) - E^{+}_{\sigma, \psi}(\mathbf{k})} \bigg] n_F[E^{-}_{\sigma, \psi}(\mathbf{k})] \bigg\rbrace, \\
Q^{B B}_{\psi, 2b} = & \; Z + \frac{U}{2 N_{c}} \sum_{\mathbf{k} \neq \mathbf{0}} \frac{\cos(k_y a)}{[E^{+}_X(\mathbf{k})]^2 - [E^{-}_X(\mathbf{k})]^2} \bigg\lbrace \bigg[ \frac{[E^{+}_X(\mathbf{k})]^2 - U E^{A A}_X(\mathbf{k})}{E^{+}_X(\mathbf{k})} \bigg] \coth\bigg[ \frac{\beta E^{+}_X(\mathbf{k})}{2} \bigg] + \bigg[ \frac{[E^{-}_X(\mathbf{k})]^2 - U E^{A A}_X(\mathbf{k})}{E^{-}_X(\mathbf{k})} \bigg] \nonumber \\
& \times \coth\bigg[ \frac{\beta E^{-}_X(\mathbf{k})}{2} \bigg] \bigg\rbrace, \\
M = & \; \frac{U}{4 N_{c}} \sum_{\sigma = \uparrow, \downarrow} \sum_{\mathbf{k}} (- 1)^{\sigma} \bigg[ \frac{E^{A A}_{\sigma, \psi}(\mathbf{k}) - E^{B B}_{\sigma, \psi}(\mathbf{k})}{E^{+}_{\sigma, \psi}(\mathbf{k}) - E^{-}_{\sigma, \psi}(\mathbf{k})} \bigg] \lbrace n_F[E^{+}_{\sigma, \psi}(\mathbf{k})] - n_F[E^{-}_{\sigma, \psi}(\mathbf{k})] \rbrace.
\end{align}

In the main text, we show the numerical solution for most of the order parameters defined by these self-consistent equations, with the exception of  $Q^{ss}_{2a, X}$, $Q^{ss}_{2b, X}$, and $\Delta_X$. The interaction dependence of the latter is shown in Fig. \ref{Fig5}. Their behavior is consistent with the discussion presented in the main text. In particular, note that $\Delta_X$ becomes nonzero at the transition from the $\mathrm{A\ell I}$ to the $\mathrm{A\ell MI}$ phase.

\section{\uppercase{Electronic spectral function}}\label{AppB}
\noindent

To demonstrate that the electron and spinon Fermi surfaces occupy the same position in momentum space, we calculate the electron spectral function 
\begin{equation}\label{Eq_SpecF}
\mathcal{A}_{\mathbf{k}}(\omega) \equiv - \frac{1}{\pi} \operatorname{Im} \lbrace \operatorname{tr} [G(\mathbf{k}, \omega + i0^{+})]\rbrace.
\end{equation}
The matrix elements of the electron Green's function on the right-hand side of Eq. \eqref{Eq_SpecF} are obtained by Fourier transforming $G^{s s^{\prime}}_{\sigma}(\boldsymbol{r} - \boldsymbol{r}^{\prime}, \tau) = - \langle T_{\tau} [ c^s_{\boldsymbol{r}, \sigma}(\tau) c^{s^{\prime} \dagger}_{\boldsymbol{r}^{\prime}, \sigma^{\prime}}(0) ] \rangle$, which in the slave-rotor approach becomes 
\begin{equation}
G^{s s^{\prime}}_{\sigma}(\boldsymbol{r} - \boldsymbol{r}^{\prime}, \tau) = - \langle T_{\tau} \lbrace e^{i [\theta^s_{\boldsymbol{r}}(\tau) - \theta^{s^{\prime}}_{\boldsymbol{r}^{\prime}}(0)]} \psi^s_{\boldsymbol{r}, \sigma}(\tau) \psi^{s^{\prime} \dagger}_{\boldsymbol{r}^{\prime}, \sigma^{\prime}}(0) \rbrace \rangle.
\end{equation} 
By means of the saddle-point approximation, one can show that the momentum-frequency Green's function $G^{s s^{\prime}}_{\sigma}(\mathbf{k}, i\omega_n)$ takes the simple form \cite{He-PRB(2022),Sangiovanni-PRL(2024)}
\begin{equation}\label{Eq_ElecProp}
G^{s s^{\prime}}_{\sigma}(\mathbf{k}, i\omega_n) = Z G^{s s^{\prime}}_{\sigma, \psi}(\mathbf{k}, i\omega_n) + \frac{T}{N_{c}} \sum_{\mathbf{q} \neq \mathbf{0}} \sum_{\nu_m} G^{s s^{\prime}}_X(\mathbf{q}, i\nu_m) G^{s s^{\prime}}_{\sigma, \psi}(\mathbf{k} - \mathbf{q}, i\omega_n - i\nu_m).
\end{equation}

We rewrite Eq. \eqref{Eq_SpecF}  as
\begin{equation}
\mathcal{A}_{\mathbf{k}}(\omega) = - \frac{1}{\pi} \sum_{\sigma = \uparrow, \downarrow} \operatorname{Im} \bigg[\sum_{s = A, B} G^{s s}_{\sigma}(\mathbf{k}, i\omega_n)\bigg\vert_{i\omega_n \rightarrow \omega + i0^{+}} \bigg].
\end{equation}
After computing the the Matsubara sum on the right-hand side of Eq. \eqref{Eq_ElecProp}, we obtain
\begin{equation}
\mathcal{A}_{\mathbf{k}}(\omega) = \sum_{\sigma = \uparrow, \downarrow} \mathcal{A}_{\mathbf{k}, \sigma}(\omega),
\end{equation} 
where
\begin{align}
\mathcal{A}_{\mathbf{k}, \sigma}(\omega) = & \; Z \sum_{s = A, B} \bigg[ \frac{\omega - E^{s s}_{\sigma, \psi}(\mathbf{k})}{E^{+}_{\sigma, \psi}(\mathbf{k}) - E^{-}_{\sigma, \psi}(\mathbf{k})} \bigg] \lbrace \delta[ \omega - E^{+}_{\sigma, \psi}(\mathbf{k})] - \delta[ \omega - E^{-}_{\sigma, \psi}(\mathbf{k})] \rbrace  \nonumber \\
&+ \frac{1}{N_{c}} \sum_{\mathbf{q} \neq \mathbf{0}} \sum_{s = A, B} [\mathcal{B}^{s}_{\sigma}(\mathbf{k}, \mathbf{q}, \omega) + \mathcal{F}^{s}_{\sigma}(\mathbf{k}, \mathbf{q}, \omega)]
\end{align}
denotes the spin-resolved spectral function. Here, $\delta(x)$ refers to the Dirac delta function, and  the two   functions appearing on the right-hand side of $\mathcal{A}_{\mathbf{k}, \sigma}(\omega)$ are defined by
\begin{align}
\mathcal{B}^{s}_{\sigma}(\mathbf{k}, \mathbf{q}, \omega) & = \frac{U \lbrace [E^{+}_X(\mathbf{q})]^2 - U E^{ss}_X(\mathbf{q}) \rbrace [ \omega - E^{s s}_{\sigma, \psi}(\mathbf{k} - \mathbf{q}) - E^{+}_X(\mathbf{q})]}{2 E^{+}_X(\mathbf{q}) \lbrace [E^{+}_X(\mathbf{q})]^2 - [E^{-}_X(\mathbf{q})]^2 \rbrace [E^{+}_{\sigma, \psi}(\mathbf{k} - \mathbf{q}) - E^{-}_{\sigma, \psi}(\mathbf{k} - \mathbf{q})]} \lbrace \delta[ E^{+}_{\sigma, \psi}(\mathbf{k} - \mathbf{q}) + E^{+}_X(\mathbf{q}) - \omega] \nonumber \\
& - \delta[ E^{-}_{\sigma, \psi}(\mathbf{k} - \mathbf{q}) + E^{+}_X(\mathbf{q}) - \omega] \rbrace n_B[E^{+}_X(\mathbf{q})] \nonumber \\
& + \frac{U \lbrace [E^{+}_X(\mathbf{q})]^2 - U E^{ss}_X(\mathbf{q}) \rbrace [ \omega - E^{s s}_{\sigma, \psi}(\mathbf{k} - \mathbf{q}) + E^{+}_X(\mathbf{q})]}{2 E^{+}_X(\mathbf{q}) \lbrace [E^{+}_X(\mathbf{q})]^2 - [E^{-}_X(\mathbf{q})]^2 \rbrace [E^{+}_{\sigma, \psi}(\mathbf{k} - \mathbf{q}) - E^{-}_{\sigma, \psi}(\mathbf{k} - \mathbf{q})]} \lbrace \delta[ E^{-}_{\sigma, \psi}(\mathbf{k} - \mathbf{q}) - E^{+}_X(\mathbf{q}) - \omega] \nonumber \\
& - \delta[ E^{+}_{\sigma, \psi}(\mathbf{k} - \mathbf{q}) - E^{+}_X(\mathbf{q}) - \omega] \rbrace n_B[- E^{+}_X(\mathbf{q})] \nonumber \\
& + \frac{U \lbrace [E^{-}_X(\mathbf{q})]^2 - U E^{ss}_X(\mathbf{q}) \rbrace [ \omega - E^{s s}_{\sigma, \psi}(\mathbf{k} - \mathbf{q}) - E^{-}_X(\mathbf{q})]}{2 E^{-}_X(\mathbf{q}) \lbrace [E^{+}_X(\mathbf{q})]^2 - [E^{-}_X(\mathbf{q})]^2 \rbrace [E^{+}_{\sigma, \psi}(\mathbf{k} - \mathbf{q}) - E^{-}_{\sigma, \psi}(\mathbf{k} - \mathbf{q})]} \lbrace \delta[ E^{-}_{\sigma, \psi}(\mathbf{k} - \mathbf{q}) + E^{-}_X(\mathbf{q}) - \omega] \nonumber \\
& - \delta[ E^{+}_{\sigma, \psi}(\mathbf{k} - \mathbf{q}) + E^{-}_X(\mathbf{q}) - \omega] \rbrace n_B[E^{-}_X(\mathbf{q})] \nonumber \\
& + \frac{U \lbrace [E^{-}_X(\mathbf{q})]^2 - U E^{ss}_X(\mathbf{q}) \rbrace [ \omega - E^{s s}_{\sigma, \psi}(\mathbf{k} - \mathbf{q}) + E^{-}_X(\mathbf{q})]}{2 E^{-}_X(\mathbf{q}) \lbrace [E^{+}_X(\mathbf{q})]^2 - [E^{-}_X(\mathbf{q})]^2 \rbrace [E^{+}_{\sigma, \psi}(\mathbf{k} - \mathbf{q}) - E^{-}_{\sigma, \psi}(\mathbf{k} - \mathbf{q})]} \lbrace \delta[ E^{+}_{\sigma, \psi}(\mathbf{k} - \mathbf{q}) - E^{-}_X(\mathbf{q}) - \omega] \nonumber \\
& - \delta[ E^{-}_{\sigma, \psi}(\mathbf{k} - \mathbf{q}) - E^{-}_X(\mathbf{q}) - \omega] \rbrace n_B[- E^{-}_X(\mathbf{q})], \\
\mathcal{F}^{s}_{\sigma}(\mathbf{k}, \mathbf{q}, \omega) & = \frac{U \lbrace [\omega - E^{+}_{\sigma, \psi}(\mathbf{k} - \mathbf{q})]^2 - U E^{ss}_X(\mathbf{q}) \rbrace [ E^{+}_{\sigma, \psi}(\mathbf{k} - \mathbf{q}) - E^{s s}_{\sigma, \psi}(\mathbf{k} - \mathbf{q})]}{2 \lbrace [E^{+}_X(\mathbf{q})]^2 - [E^{-}_X(\mathbf{q})]^2 \rbrace [E^{+}_{\sigma, \psi}(\mathbf{k} - \mathbf{q}) - E^{-}_{\sigma, \psi}(\mathbf{k} - \mathbf{q})]} \bigg( \frac{1}{E^{+}_X(\mathbf{q})} \lbrace \delta[ \omega - E^{+}_{\sigma, \psi}(\mathbf{k} - \mathbf{q}) - E^{+}_X(\mathbf{q})] \nonumber \\
& - \delta[ \omega - E^{+}_{\sigma, \psi}(\mathbf{k} - \mathbf{q}) + E^{+}_X(\mathbf{q})] \rbrace + \frac{1}{E^{-}_X(\mathbf{q})} \lbrace \delta[ \omega - E^{+}_{\sigma, \psi}(\mathbf{k} - \mathbf{q}) + E^{-}_X(\mathbf{q})] - \delta[ \omega - E^{+}_{\sigma, \psi}(\mathbf{k} - \mathbf{q}) - E^{-}_X(\mathbf{q})] \rbrace \bigg) \nonumber \\
& \times n_F[ - E^{+}_{\sigma, \psi}(\mathbf{k} - \mathbf{q})] \nonumber \\
& + \frac{U \lbrace [\omega - E^{-}_{\sigma, \psi}(\mathbf{k} - \mathbf{q})]^2 - U E^{ss}_X(\mathbf{q}) \rbrace [ E^{-}_{\sigma, \psi}(\mathbf{k} - \mathbf{q}) - E^{s s}_{\sigma, \psi}(\mathbf{k} - \mathbf{q})]}{2 \lbrace [E^{+}_X(\mathbf{q})]^2 - [E^{-}_X(\mathbf{q})]^2 \rbrace [E^{+}_{\sigma, \psi}(\mathbf{k} - \mathbf{q}) - E^{-}_{\sigma, \psi}(\mathbf{k} - \mathbf{q})]} \bigg( \frac{1}{E^{+}_X(\mathbf{q})} \lbrace \delta[ \omega - E^{-}_{\sigma, \psi}(\mathbf{k} - \mathbf{q}) + E^{+}_X(\mathbf{q})] \nonumber \\
& - \delta[ \omega - E^{-}_{\sigma, \psi}(\mathbf{k} - \mathbf{q}) - E^{+}_X(\mathbf{q})] \rbrace + \frac{1}{E^{-}_X(\mathbf{q})} \lbrace \delta[ \omega - E^{-}_{\sigma, \psi}(\mathbf{k} - \mathbf{q}) - E^{-}_X(\mathbf{q})] - \delta[ \omega - E^{-}_{\sigma, \psi}(\mathbf{k} - \mathbf{q}) + E^{-}_X(\mathbf{q})] \rbrace \bigg) \nonumber \\
& \times n_F[ - E^{-}_{\sigma, \psi}(\mathbf{k} - \mathbf{q})],
\end{align}
where $n_B(z) = (e^{\beta z} - 1)^{- 1}$ is the Bose-Einstein distribution function.

\twocolumngrid


%

\onecolumngrid

\end{document}